\documentclass{llncs}

\usepackage[british]{babel}
\usepackage[utf8]{inputenc}
\usepackage{listings}
\usepackage{Common/prooftree}
\usepackage{lmodern}
\usepackage{amsmath,amssymb,stmaryrd}
\usepackage[a4paper=false,colorlinks,pagebackref,linkcolor=blue,citecolor=red,urlcolor=blue]{hyperref}
\usepackage{xfrac}
\usepackage{upgreek}
\usepackage{wrapfig}
\usepackage{hyperref}
\usepackage{Common/appendix}

\capturecounter{theorem}
\capturecounter{lemma}
\capturecounter{definition}
\capturecounter{figure}

\newcommand\longversion[1]{#1}
\newcommand\shortversion[1]{\ignore{#1}}

\pdfpagesattr{/CropBox [77 62 538 768]}

\long\def\ignore#1{\relax}

\newcommand\struto[1][15pt]{{\raise #1 \hbox{\strut}}}%
\newcommand\strutb[1][15pt]{{\raise-#1 \hbox{\strut}}}%

\newcommand\upline{\hline\struto[5pt]}
\newcommand\midline[1][5pt]{\\[#1]\hline\struto[5pt]}

\newcommand\downline[1][5pt]{\\[#1]\hline}

\makeatletter
\def\@boxfigurewith[#1]{\figure[#1]\vbox\bgroup\hrule height.1em}
\def\@boxfigurewithout{\figure\vbox\bgroup\hrule height.1em}

\makeatother

 \newcommand\toaux[1]{\immediate\write\@auxout{#1}}

\newcommand\oldparforMain{}
\let\oldparforMain\par
\newcommand\topequationskip{.2\baselineskip}
\newcommand\botequationskip{.2\baselineskip}

\makeatletter
\renewcommand\[[1][\topequationskip]{\begingroup\let\mysavedpar\par\let\par\oldparforMain\vskip#1\nopagebreak\hbox to\hsize\bgroup\hfil\(}
\renewcommand\][1][\botequationskip]{\)\hfil\egroup\hrule height 0pt\endgroup\@afterindentfalse%\@afterheading
  \vskip#1\noindent}
\makeatother

\newbox\columnsbox
\newbox\tmpbox
\newdimen\columnsheight
\newdimen\columnwidth
\newdimen\remainingwidth
\newdimen\textwidthsave
\def\mycolumnsheight{}

\newcommand\columns[1]{%
  \def\mycolumnsheight{}%
  \setlength\remainingwidth\textwidth%
  \setbox\columnsbox=\vbox\bgroup\vskip0pt\vfil\hbox to\textwidth\bgroup#1\egroup\vfil\egroup%
  \columnsheight=\ht\columnsbox%
  \def\mycolumnsheight{to\columnsheight}%
  \hrule height 0pt\vtop{\hbox to\wd\columnsbox\bgroup#1\egroup}%
}

\makeatletter

\def\commonpart{%
  \setlength\columnwidth{\wd\tmpbox}%
  \vtop{\vskip0pt\hbox to\columnwidth{{\box\tmpbox}}}%
  \advance\remainingwidth-\columnwidth%
  \setlength\textwidth\textwidthsave%
  \hsize\textwidthsave%
}
\def\column{\unskip\setlength\textwidthsave\textwidth\@ifnextchar[\@columnwith\@columnwithout}
\long\def\@columnwith[#1]#2{%
  \def\newhsize{#1\dimexpr\textwidth\relax}%
  \hsize\newhsize%
  \ifdim\hsize<0.1pt\hsize\remainingwidth\fi%
  \setlength\textwidth\hsize%
  \setbox\tmpbox=\hbox to\hsize\bgroup\hfil\vtop\mycolumnsheight{\vskip0pt#2\vskip0pt}\hfil\egroup%
  \commonpart%
}
\long\def\@columnwithout#1{%
  \hsize\remainingwidth%
  \setlength\textwidth\hsize%
  \setbox\tmpbox=\hbox\bgroup\vtop\mycolumnsheight{\vskip0pt#1\vskip0pt}\egroup%
  \commonpart%
}
\makeatother

\newenvironment{centre}{\begin{center}\unskip}{\end{center}\unskip}

\newcommand\vect[1]{{\overrightarrow{#1}}}            % vecteur
\renewcommand{\iff}{if and only if}

\newcommand{\ie}{i.e.~}
\newcommand{\eg}{e.g.~}

\newcommand{\wrt}{w.r.t.~}
\newcommand{\resp}{resp.~}

\newcommand{\eqdef}{:=\ }

\newcommand{\sqin}{\textsf{\footnotesize{E}}}

\newcommand{\Gam}{\Gamma}

\newcommand{\Del}{\Delta}

\newcommand{\sig}{\sigma}

\newcommand\mathFomega{F_\omega}
\newcommand\Fomega{\ifmmode\mathFomega\else$\mathFomega$\fi}
\newcommand\mathFomegaC{F_\omega^{\mathcal C}}
\newcommand\FomegaC{\ifmmode\mathFomegaC\else$\mathFomegaC$\fi}
\newcommand\mathDNE{\mathrm{DNE}}
\newcommand\DNE{\ifmmode\mathDNE\else$\mathDNE$\fi}

\newcommand\Psyche{\textsc{Psyche}}

\newcommand\FV[2][{}]{\textsf{FV}_{#1}(#2)}

\newcommand{\seqg}[3]{\mbox{$\ {#1}_{#2}^{#3}\ $}}
\newcommand{\seqf}[2][]{\seqg{\vdash}{#1}{#2}}

\newcommand\DerPos[3]{{#1}   \seqf{}   {[#2]}}
\newcommand\DerNeg[3]{{#1}   \seqf{}   {#2}}

\newcommand\daggerL{\raise3pt\hbox{\rotatebox{-40}{$\dagger$}}}
\newcommand\daggerR{\raise0pt\hbox{\rotatebox{40}{$\dagger$}}}

\newcommand\andN{{\wedge}}
\newcommand\orP{{\vee}}

\newcommand\EX[2]{\exists #1 #2}
\newcommand\FA[2]{\forall #1 #2}

\newcommand{\non}[1]{\overline{#1}}

\newcommand\LKThp[1][\mathcal T]{\textsf{LK}$^p$($#1$)}

\newcommand\BB[1]{\models\!#1}
\newcommand\BBc[4][]{{#1}\models^{#4}\!#2\gives #3}
\newcommand\BBr[4]{\ensuremath{{#1}\gives\models\hskip-.3em\rangle^{#4}\!#2\gives #3}}
\newcommand\BBi{\ensuremath{\models}}
\newcommand\BBci[1]{\ensuremath{\models^{#1}}}
\newcommand\BBri[1]{\ensuremath{\models\hskip-.3em\rangle^{#1}}}

\newcommand\LKone{{\(\sf LK_{1}\)}}
\newcommand\DI{{\(\sf LK^?_{1}\)}}
\newcommand\SDI{{\(\sf LK^{?\hskip-.03em{\rangle}}_{1}\)}}

\newcommand\DPLLTh{\textsf{DPLL}($\mathcal T$)}

\newcommand\atmCtxt[1]{{#1}_{\textsf{lit}}}

\newcommand\Init[1][]{\textsf{Init}_{#1}}
\newcommand\Release{\textsf{Release}}
\newcommand\Select{\textsf{Select}}
\newcommand\Store{\textsf{Store}}

\renewcommand\sqin{\raise1pt\hbox{\large $\,\epsilon\,$}}

\newcommand\Index[2][]{\emph{#2}}

\newcommand{\meet}{\hskip-0.15em\wedge\hskip-0.15em}
\newcommand{\compat}{\upepsilon}

\newcommand{\gives}{\shortrightarrow}
\newcommand{\into}{\shortrightarrow}

\newcommand\XVar[1]{\ensuremath{?\hspace{-.1em}{#1}}}
\newcommand\EVar[1]{\ensuremath{{\textsc{#1}}}}

\newcommand\inject[1]{#1^{\uparrow}}
\newcommand\project[1]{#1_{\downarrow}}
\newcommand\bind[1]{f_{#1}}
\newcommand\fold[1]{\textsf{fold}{#1}}
\newcommand\stronger[1][]{\leqslant_{#1}}
\newcommand\obseq[1][]{\simeq_{#1}}
\newcommand\addFV[2]{#1;#2}

\newcommand\eqlit[2]{#1\hspace{-.3em}=\hspace{-.3em}#2}
\newcommand\neqlit[2]{#1\hspace{-.3em}\neq\hspace{-.3em}#2}

\newcommand\dinit{d_0}

\newcommand\Inst[1]{\ensuremath{\Sigma_{#1}}}
\newcommand\addInst[3]{\ensuremath{#3,\hspace{-.2em}#1\hspace{-.2em}\mapsto\hspace{-.2em}#2}}
\newcommand\emptyInst{\emptyset}

\newcommand\AXproj[1][]{\ensuremath{\textsf{Proj}}}
\newcommand\AXmeet{\ensuremath{\textsf{Meet}}}
\newcommand\AXwitness{\ensuremath{\textsf{Wit}}}
\newcommand\AXpg{\ensuremath{\textsf{PG}}}

\newcommand\AXlift[1][]{\ensuremath{\textsf{Lift}}}
\newcommand\AXp[1][]{\ensuremath{P_{#1}}}
\newcommand\AXrp{\ensuremath{\textsf{R2P}}}
\newcommand\AXpr{\ensuremath{\textsf{P2R}}}

\begin{document}

\title{Axiomatic constraint systems for proof search modulo
  theories
  % \thanks{This research was supported by ANR projects PSI and
  %   ALCOCLAN, and by DARPA under agreement number
  %   FA8750-12-C-0284. The views and conclusions contained herein are
  %   those of the authors and should not be interpreted as necessarily
  %   representing the official policies or endorsements, either
  %   expressed or implied, of DARPA, or the U.S.\ Government.  }
}

\author{
    Damien Rouhling\inst{1} \and
    Mahfuza Farooque\inst{2} \and\\
    Stéphane Graham-Lengrand\inst{2,3,4} \and
    Assia Mahboubi\inst{3} \and
    Jean-Marc Notin\inst{2}
}

\authorrunning{D. Rouhling, M. Farooque, S. Graham-Lengrand, A.
Mahboubi and J.-M. Notin}

\institute{
    École Normale Supérieure de Lyon, France \and
    CNRS - École Polytechnique, France \and
    INRIA, Centre de Recherche en Informatique Saclay-Île de France \and
    SRI International, USA
}

\maketitle

\begin{abstract}
  Goal-directed proof search in first-order logic uses meta-\linebreak
  variables to delay the choice of witnesses; substitutions for such
  variables are produced when closing proof-tree branches, using
  first-order unification or a theory-specific background
  reasoner. This paper investigates a generalisation of such
  mechanisms whereby \emph{theory-specific constraints} are produced
  instead of substitutions. In order to design modular proof-search
  procedures over such mechanisms, we provide a sequent calculus with
  meta-variables, which manipulates such constraints
  abstractly. Proving soundness and completeness of the calculus leads
  to an axiomatisation that identifies the conditions under which
  abstract constraints can be generated and propagated in the same way
  unifiers usually are. We then extract from our abstract framework a
  component interface and a specification for concrete implementations
  of background reasoners.
\end{abstract}

\section{Introduction}\label{sec:intro}

A broad literature studies the integration of theory reasoning with
generic automated reasoning techniques. Following Stickel's seminal
work~\cite{Stickel85}, different levels of interaction have been
identified~\cite{baumgartner1992unified} between a theory-generic
\emph{foreground reasoner} and a theory-specific \emph{background
  reasoner}, with a specific scheme for the \emph{literal level} of
interaction. In absence of quantifiers, the \DPLLTh\
architecture~\cite{Nieuwenhuis06} is an instance of the scheme and a
successful basis for SMT-solving, combining SAT-solving techniques for
boolean logic with a procedure that decides whether a conjunction of
ground literals is consistent with a background theory $\mathcal T$.

Our contribution falls into such a scheme, but in presence of
quantifiers, and hence of non-ground literals. When given a
conjunction of these, the background reasoner provides a means to make
this conjunction inconsistent with $\mathcal T$, possibly by
instantiating some
(meta-)variables~\cite{baumgartner1992unified}. Technically, it
produces a \emph{$\mathcal T$-refuter} that contains a substitution.

Beckert~\cite{beckert99handbook} describes how this approach can be
applied to \emph{analytic tableaux}, in particular \emph{free variable
  tableaux}: $\mathcal T$-refuters are produced to extend and
eventually close a tableau branch, while the substitutions that they
contain are globally applied to the tableau, thus affecting the
remaining open branches. In fact, the \emph{only} way in which closing
a branch affects the other branches is the propagation of these
substitutions, as it is the case for tableaux without theory
reasoning.  This is well-suited for some theories like equality, for
which \emph{rigid E-unification} provides a background reasoner (see
\eg\cite{Beckert98}), but maybe not for other theories. For instance,
the case of Linear Integer Arithmetic (LIA) was addressed by using
arithmetic constraints, and quantifier elimination, in the Model
Evolution calculus~\cite{BFT2008MELIA} and the Sequent
Calculus~\cite{princess08} (which is closer to the above tableaux).

% But, as explained in
% \eg\cite{BFT2008MELIA,princess08} and in Sect.~\ref{sec:background}, this may
% not be so well-suited for other theories such as Linear Integer
% Arithmetic
% (LIA). % \footnote{Imagine that a branch may be closed by instantiating a
%   % meta-variable $\XVar X$ by any positive integer; if the only way to
%   % impact the rest of the problem is substitution, one must blindly
%   % commit to a particular integer, which may later prove a poor
%   % choice.}
% The issue for LIA was addressed by using arithmetic
% constraints, and quantifier elimination, in the Model Evolution
% calculus for~\cite{BFT2008MELIA} and the Sequent Calculus
% for~\cite{princess08} % \footnote{For instance in the sequent calculus
%   % of~\cite{princess08}, closing a branch generates a constraint such
%   % as $0\leq\XVar X$ that is further propagated along the proof-tree.}
% (which is closer to the free variable tableaux discussed above).

This paper develops sequent calculi with a more general
\emph{abstract} notion of constraints so that more theories can be
treated in a similar way, starting with all theories admitting quantifier
elimination. But it also covers those total theories (total
in the sense that $\mathcal T$-refuters are just substitutions)
considered by Beckert~\cite{beckert99handbook} for free variable
tableaux, for which constraints are simply substitutions.

Sect.~\ref{sec:background} presents a sequent calculus \LKone\ with
ground theory reasoning (as in \DPLLTh) and various target theories
that we intend to capture. Sect.~\ref{sec:conssys} introduces our
abstract systems of constraints. Sect.~\ref{sec:delayed} presents a
sequent calculus \DI\ similar to Rümmer's PresPred${}^C_S$
calculus~\cite{princess08}, but generalised with abstract
constraints. It collects constraints from the parallel/independent
exploration of branches, with the hope that their combination remains
satisfiable. Sect.~\ref{sec:sequentialising} and~\ref{sec:relating}
present a variant \SDI\ where the treatment of branching is
asymmetric, reflecting a sequential implementation of proof search:
the constraint that is produced to close one branch
%  is given as an
% extra input for 
affects the exploration of the next branch, 
% which is much
% closer to
as in free variable tableaux~\cite{beckert99handbook}. Each time,
we prove soundness and completeness relative to the reference sequent
calculus \LKone. From these proofs we extract an axiomatisation for
our background theory reasoner and its associated constraints. In
Sect.~\ref{sec:implem} this axiomatisation is used to
define a component interface with a formal specification, for our
quantifier-handling version 2.0 of the \Psyche\ platform for theorem
proving~\cite{Psyche}.  We conclude by discussing related works and
future work.

\section{Ground calculus and examples}\label{sec:backexamples}

\label{sec:background}

The simple sequent calculus that we use in this paper uses the
standard first-order notions of term, literal, eigenvariable, and
formula. Following standard practice in \emph{tableaux} methods or the
linear logic tradition, we opt for a compact one-sided presentation of
the sequent calculus, here called \LKone. Its rules are presented in
Fig.~\ref{sys1}, where $\Gamma$ is a set (intuitively seen as a
disjunction) of first-order formulae (in negation-normal form) and
$\atmCtxt\Gamma$ is the subset of its literals; $A[x:=t]$ denotes the
substitution of term $t$ for all free occurrences of variable $x$ in
formula $A$; finally, $\models$ denotes a specific predicate, called
the \emph{ground validity predicate}, on sets of \emph{ground}
literals (\ie literals whose variables are all eigenvariables). This
predicate is used to model a given theory $\mathcal T$, with the
intuition that $\BB{\atmCtxt\Gamma}$ holds when the disjunction of the
literals in $\Gamma$ is $\mathcal T$-valid. Equivalently, it holds
when the conjunction of their negations is $\mathcal T$-inconsistent,
as checked by the decision procedures used in SMT-solving. Likewise,
checking whether $\BB{\atmCtxt\Gamma}$ holds is performed by a
background reasoner, while the bottom-up application of the rule of
\LKone\ can serve as the basis for a \emph{tableaux}-like foreground
reasoner.\pagebreak

\begin{figure}[!h]
  \[
  \boxed{
    \begin{array}{c}
      \infer[\BB{\atmCtxt\Gamma}]{\vdash\Gamma}{}
      \qquad
      \infer{\vdash\Gamma,A\wedge B}{\vdash\Gamma,A \qquad \vdash\Gamma,B}
      \qquad
      \infer{\vdash\Gamma,A\vee B}{\vdash\Gamma,A,B}
      \\[15pt]
      \begin{array}{c}
        \infer{\vdash\Gamma,\exists x A}{\vdash\Gamma,A\left[x:=t\right],\exists x A}\\
        \strut
      \end{array}
      \qquad
      \begin{array}{c}
        \infer{\vdash\Gamma,\forall x A}{\vdash\Gamma,A\left[x:=\EVar{x}\right]}\\
        \mbox{\scriptsize where $\EVar{x}$ is a fresh eigenvariable}
      \end{array}
    \end{array}
  }
  \]
  \caption{The \LKone\ sequent calculus modulo theories}
  \label{sys1}
\end{figure}

But a realistic proof-search procedure is in general unable to provide
an appropriate witness $t$ ``out of the blue'' at the time of applying
an existential rule. We shall use \emph{meta-variables} (called
\emph{free variables} in tableaux terminology) to delay the production
of such instances until the constraints of completing/closing branches
impact our choice possibilities. The way this happens heavily depends
on the background theory, and below we give a few examples (more
background on the technical notions can be found for instance in
Beckert's survey~\cite{beckert99handbook}):

\begin{example}[Pure first-order logic]\label{ex:pure1st-order}
  In the empty theory, closing a branch $\vdash\Gamma$ is done by
  finding a literal $l$ and its negation $\non l$ in $\Gamma$ or, if
  meta-variables were used, by finding a pair $l$ and $\non {l'}$ and
  a substitution $\sigma$ for meta-variables that \emph{unifies} $l$
  and $l'$. Such a \emph{first-order unifier} $\sigma$ may be produced
  by the sole analysis of $\vdash\Gamma$, or by the simultaneous analysis of
  the branches that need to be closed. Since the latter problem is
  still decidable, a global management of unification
  constraints is sometimes preferred, avoiding the propagation of
  unifiers from branch to branch.
\end{example}

\begin{example}[First-order logic with equality]\label{ex:eq}
  When adding equality, closing a branch $\vdash\Gamma$ is done by
  finding in $\Gamma$ either an equality $\eqlit t u$ such that
  $\Gamma_E\models_E \eqlit t u$, or a pair of literals
  $p(t_1,\ldots,t_n)$ and $\non {p}(u_1,\ldots,u_n)$ such that
  $\Gamma_E\models_E \eqlit{t_1}{u_1}\wedge\cdots\wedge
  \eqlit{t_n}{u_n}$, where $\Gamma_E$ is the set of all equalities
  $\eqlit a b$ such that $\neqlit a b$ is in $\Gamma$, and $\models_E$
  is entailment in the theory of equality. \emph{Congruence closure}
  can be used to check this entailment.  If meta-variables were used,
  then a substitution $\sigma$ for meta-variables has to be found such
  that \eg $\sigma(\Gamma_E)\models_E \eqlit{\sigma(t)}{\sigma(u)}$, a
  problem known as \emph{rigid E-unification}. While this problem is
  decidable, finding a substitution that simultaneously closes several
  open branches (\emph{simultaneous rigid E-unification}) is
  undecidable. A natural way to use %a decision procedure for
  rigid E-unification is to produce a \emph{stream} of substitutions
  from the analysis of one branch and propagate them into the other
  branches; if at some point we have difficulties closing one of
  these, we can try the next substitution in the stream.
\end{example}

The idea of producing streams of substitutions at the leaves of branches
(advocated by Giese~\cite{Giese00}) can be taken further:

\begin{example}[Theories with ground decidability]\label{ex:ground}
  Any theory whose ground validity predicate is decidable has a
  semi-decision procedure that ``handles''\linebreak meta-variables: to close a
  branch $\vdash\Gamma$ with meta-variables, enumerate as a stream all
  substitutions to ground terms (\ie terms whose variables are all
  eigenvariables), and filter out of it all substitutions $\sigma$
  such that $\not\models\atmCtxt{\sigma(\Gamma)}$. Stream productivity
  -and therefore decidability- may thus be lost, but completeness of
  proof search in first-order logic already requires fairness of
  strategies with \eg iterative deepening methods, which may as well
  include the computation of streams.

  While this mostly seems an impractical theoretical remark,
  heuristics can be used (\eg first trying those ground terms that are
  already present in the problem) that are not far from what is
  implemented in SMT-solvers (like \emph{triggers}~\cite{DrossCKP12}).

  The enumeration strategy can also be theory-driven, and also make
  use of substitutions to non-ground terms: An interesting instance of
  this is higher-order logic expressed as a first-order theory,
  $\lambda$-terms being encoded as first-order terms using De Bruijn's
  indices, and $\beta\eta$-equivalence being expressed with
  first-order axioms. Similarly to Example~\ref{ex:pure1st-order},
  closing a branch with meta-variables requires solving (higher-order)
  unification problems, whose (semi-decision) algorithms can be seen
  as complete but optimised enumeration techniques.
\end{example}

All of the above examples use substitutions of meta-variables as the
output of a successful branch closure, forming \emph{total background
  reasoners} for the tableaux of~\cite{beckert99handbook}. But by
letting successful branch closures produce a more general notion of
theory-specific \emph{constraints}, we also cover examples such as:

\begin{example}[Theories with quantifier elimination]\label{ex:qe}
  When a theory satisfies quantifier elimination (such as linear
  arithmetic), the provability of arbitrary formulae can be reduced to
  the provability of quantifier-free formulae. This reduction can be
  done with the same proof-search methodology as for the previous
  examples, provided successful branch closures produce other kinds of
  data-structures. For instance with $p$ an uninterpreted predicate
  symbol, $l(x,y):=3x\leq 2y\leq 3x\!+\!1$ and $l'(x,y):=99\leq
  3y\!+\!2x\leq 101$, the foreground reasoner will turn the sequent
  \[\vdash(\exists xy(p(x,y)\wedge l(x,y)))\vee(\exists x'y'(\non p(x',y')\wedge l'(x',y'))\]
  into a tree with 4 branches, with meta-variables $\XVar X$, $\XVar {X'}$, $\XVar Y$, and $\XVar {Y'}$:
  \[\begin{array}{cc}
    \vdash p(\XVar X,\XVar Y),\non p(\XVar {X'},\XVar {Y'})
    &\quad
    \vdash l(\XVar X,\XVar Y),\non p(\XVar {X'},\XVar {Y'})\\
    \vdash p(\XVar X,\XVar Y),l'(\XVar {X'},\XVar {Y'})
    &\quad
    \vdash l(\XVar X,\XVar Y),l'(\XVar {X'},\XVar {Y'})
  \end{array}
  \]
  While it is clear that the background reasoner will close the
  top-left leaf by producing the substitution identifying $\XVar X$
  with $\XVar {X'}$, $\XVar Y$ with $\XVar {Y'}$, it is hard to see
  how the analysis of any of the other branches could produce, on its
  own and not after a lengthy enumeration, a substitution that is, or
  may be refined into, the unique integer solution $\XVar X\mapsto 15,
  \XVar Y\mapsto 23$.  Hence the need for branches to communicate to
  other branches more appropriate data-structures than substitutions,
  like \emph{constraints} (in this case, arithmetic ones).
\end{example}

In the rest of this paper, all of the above examples are instances of
an abstract notion of theory module that comes with its own system of constraints.

\section{Constraint Structures}
\label{sec:conssys}
% In this section we introduce and discuss the constraint structures
% used in the calculi presented in Sect.~\ref{sec:delayed}
% and~\ref{sec:sequentialising}.

Meta-variables (denoted $\XVar X$, $\XVar Y$, etc) can be thought of
as place-holders for yet-to-come instantiations. Delayed though these
may be, they must respect the freshness conditions from System \LKone,
so dependencies between meta-variables and eigenvariables must be
recorded during proof search.

While Skolem symbols are a convenient implementation of such
dependencies when the theory reasoner is unification-based
(\emph{occurs check} ruling out incorrect instantiations for free), we
record them in a data-structure, called \emph{domain}, attached to
each sequent.  Two operations are used on domains: adding to a domain
$d$ a fresh eigenvariable $\EVar x$ (\resp meta-variable $\XVar X$)
results in a new domain $\addFV d {\EVar x}$ (\resp $\addFV d {\XVar
  X}$). The use of the notation always implicitly assumes $\EVar x$
(\resp $\XVar X$) to be fresh for $d$. An \emph{initial domain}
$\dinit$ is also used before proof search introduces fresh
eigenvariables and meta-variables.\footnote{For instance, a domain may
  be implemented as a pair $(\Phi;\Delta)$, where $\Phi$ is the set of
  declared eigenvariables and $\Delta$ maps every declared
  meta-variable to the set of eigenvariables on which it is authorised
  to depend.  With this implementation, $\addFV {(\Phi;\Delta)} {\EVar
    x}\eqdef (\Phi,\EVar x;\Delta)$ and $\addFV {(\Phi;\Delta)} {\XVar
    X}\eqdef (\Phi;\Delta,\XVar X\mapsto\Phi)$.  We also set
  $\dinit=(\Phi_0,\emptyset)$, with $\Phi_0$ already containing enough
  eigenvariables so as to prove \eg $\exists x (p(x)\vee \non p(x))$.}

\begin{definition}[Terms, formulae with meta-variables]
  A \emph{term} (\resp \emph{formula}) \emph{of domain $d$} is a term (\resp
  formula) whose variables (\resp free variables) are all
  eigenvariables or meta-variables declared in $d$.
  A term (\resp formula) is \emph{ground} if it contains no meta-variables.
  % For $n\in \mathbb{N}$, let $T_{n}$ be the set of ground terms whose
  % eigenvariables have indexes below $n$.
  Given a domain $d$, we define $T_d$ to be the set of ground terms of domain
  $d$.
  A \emph{context of domain $d$} is a multiset of
  formulae of domain $d$. In the rest of this paper, a \emph{free
    variable} (of domain $d$) means either an eigenvariable or a
  meta-variable (declared in $d$).
  % A term (\resp formula, context) of domain $l$ is a term (\resp
  % formula, context) with meta-variables among $\XVar
  % 1,\ldots,\XVar{\abs l}$ (besides bound variables and
  % eigenvariables). The arity of the meta-variable $\XVar n$ in domain
  % $l$ is the $(\abs l - n +1)^{th}$ element of $l$. A term of domain
  % $\emptylist$ (i.e. without meta-variable) is called a \emph{ground term}.
\end{definition}

% \begin{remark}\label{rmk:prefix}  Any term $u$ (\resp formula,
%   context) on a domain $d_1$ also has domain $d_2$ if $d_2$ extends $d_1$.
% %% % and for domain $d_2$ extending $d_1$ the
% %% %   term $u$ (\resp formula, context) also has domain $d_2$.
% %% % For any term $u$ (\resp formula,
% %% %   context) on a domain $d_1$ and for domain $d_2$ extending $d_1$ the
% %% %   term $u$ (\resp formula, context) also has domain $d_2$.
% \end{remark}

In this setting, the axiom rule of system \LKone\ is adapted to the
presence of meta-variables in literals, so as to produce
theory-specific \emph{constraints} on (yet-to-come) instantiations. We
characterise the abstract structure that they form:

% , whose satisfaction
% is \emph{sufficient} to close the branch.

% When possible, generating a necessary and sufficient constraint would
% be the best option, but we do not enforce this. We then propagate
% those constraints between branches and combine them, backtracking
% whenever the carried constraint becomes unsatisfiable, but hopefully
% closing all branches.

% An abstract structure of constraints is therefore formalized as follows:
\begin{definition}[Constraint structures]
  A \Index{constraint structure} is:
  \begin{itemize}
  \item a family of sets $(\Psi_{d})_d$, indexed by domains and
    satisfying $\Psi_{\addFV d{\EVar x}}=\Psi_d$ for all domains $d$
    and eigenvariables ${\EVar x}$; elements of $\Psi_{d}$ are called
    \Index[constraint]{constraints of domain} $d$, and are denoted
    $\sigma, \sigma',$ etc.
  \item a family of mappings from $\Psi_{\addFV d {\XVar X}}$ to
    $\Psi_{d}$ for all domains $d$ and meta-variables $\XVar X$,
    called \emph{projections}, mapping constraints
    $\sigma\in\Psi_{\addFV d {\XVar X}}$ to constraints $\project
    \sigma\in\Psi_{d}$.
  \end{itemize}

  A \Index{meet constraint structure} is a constraint structure $(\Psi_{d})_d$
  with a binary operator $(\sig,\sigma')\mapsto \sig\meet\sig'$ on each set $\Psi_d$.

  A \Index{lift constraint structure} is a constraint structure
  $(\Psi_{d})_d$ with a map $\sig\mapsto \inject\sig$ from $\Psi_d$ to
  $\Psi_{\addFV d {\XVar X}}$ for all domains $d$ and meta-variables ${\XVar X}$.
%%  Strict monotonicity means that
%%  \[\begin{array}{ll}
%%    \textsf{M}&\forall\sig,\sig',\ \sig\stronger\sig'\Rightarrow\inject\sig\stronger\tilde\sig'\\
%%    \textsf{S}&\inject\bot = \bot
%%  \end{array}
%%  \]
\end{definition}

Intuitively, each mapping from $\Psi_{\addFV d {\XVar X}}$ to
$\Psi_{d}$ projects a constraint concerning the meta-variables
declared in $(\addFV d {\XVar X})$ to a constraint on the meta-variables in $d$.
Different constraints can be used for different theories:
\begin{example}\label{ex:constraintstruct}
  \begin{enumerate}
  \item In Examples~\ref{ex:pure1st-order} and~\ref{ex:eq}, it is
    natural to take $\Psi_d$ to be the set whose elements are either
    $\bot$ (to represent the unsatisfiable constraint) or a
    substitution $\sigma$ for the meta-variables in
    $d$.\footnote{Technically, the term $\sigma(\XVar X)$, if defined,
      features only eigenvariables among those authorised for
      $\XVar X$ by $d$, and meta-variables outside $d$ or mapped to
      themselves by $\sig$.} Projecting a substitution
    from $\Psi_{\addFV d {\XVar X}}$ is just erasing its entry for
    $\XVar X$. The meet of two substitutions is their most general
    unifier, and the lift of $\sigma \in \Psi_{d}$ into
    $\Psi_{\addFV d {\XVar X}}$ is $\addInst{\XVar{X}}{\XVar{X}}\sig$.
  \item In Example~\ref{ex:ground}, the default constraint structure
    would restrict the above to substitutions that map meta-variables
    to either themselves or to ground terms, unless a particular
    theory-specific enumeration mechanism could make use of non-ground
    terms (such as higher-order unification).
  \item In Example~\ref{ex:qe}, $\Psi_{\addFV d{\EVar x}}=\Psi_d$ for
    any $d$, and we take $\Psi_{\dinit}$ (\resp
    $\Psi_{\addFV d {\XVar X}}$) to be the set of quantifier-free
    formulae of domain $\dinit$ (\resp $\addFV d {\XVar X}$).
    Quantifier elimination provides
    projections, the meet operator is conjunction and the lift is
    identity.
    % We take
    % $\Psi_l$ as the  set of quantifier-free formulae with domain $l$ and
    % whose eigenvariables are among those on which some meta-variable may
    % depend according to $l$.
  \end{enumerate}
\end{example}

\section{A system for proof search with constraints}
\label{sec:delayed}

% In this section we adapt the sequent calculus \LKone\ in order to delay
% instantiations, using meta-variables and a meet constraint structure.

In the rest of this section $(\Psi_{d})_d$ denotes a fixed meet constraint structure.

\subsection{The constraint-producing sequent calculus \DI}

This sequent calculus is parameterised by a background theory reasoner
that can handle meta-variables. The reasoner is modelled by a
\emph{constraint-producing predicate} that generalises the ground
validity predicate used in System~\LKone{}.

% The black box of system \LKone, which checked the validity (in the
% background theory) of a disjunction of literals, now has to handle
% meta-variables and therefore has more work to do: it takes as input a
% domain $l$ and a set of literals of domain $l$, and outputs a
% constraint in $\Psi_{l}$. Intuitively, the constraint represents a
% sufficient condition on the meta-variables for the disjunction of
% literals to be valid in the theory. This more advanced black box is
% formalised as a \emph{constraint-producing predicate}, from which the
% sequent calculus \DI\ is defined:

\begin{definition}[\DI\ sequent calculus]

  A \emph{constraint-producing predicate} is a family of relations
  $(\models^d)_d$, indexed by domains $d$, relating sets $\mathcal A$ of
  literals of domain $d$ with constraints $\sigma$ in $\Psi_d$; when
  it holds, we write $\BBc{\mathcal A}\sig d$.

  Given such a predicate $(\models^d)_d$, the
  \Index{constraint-producing sequent calculus} \DI\ manipulates
  sequents of the form $\vdash^{d}\Gamma\gives\sig$, where $\Gamma$ is
  a context and $\sigma$ is a constraint, both of domain $d$. Its
  rules are presented in Fig.~\ref{sys2}.
\end{definition}

\begin{figure}[h]
  \[
  \boxed{\begin{array}{c}
    %% Axiom_R
    \infer[\BBc{\atmCtxt\Gamma}\sig d]{\vdash^{d}\Gamma\gives\sig}{}
    \qquad
    %% And_R
    \infer{\vdash^{d}\Gamma,A\wedge B\gives\sig\meet\sig'}{\vdash^{d}\Gamma,A\gives\sig \qquad \vdash^{d}\Gamma,B\gives\sig'}
    \qquad
    %% Or_R
    \infer{\vdash^{d}\Gamma,A\vee B\gives\sig}{\vdash^{d}\Gamma,A,B\gives\sig}
    \\\\
    %% Exists_R
    \begin{array}{c}
    \infer{\vdash^{d}\Gamma,\exists x A\gives\project\sig}{\vdash^{\addFV{d}{\XVar{X}}}\Gamma,A\left[x:=\XVar{X}\right],\exists x A\gives\sig}\\
    \mbox{\scriptsize where $\XVar{X}$ is a fresh meta-variable}\\
    \end{array}
    \qquad
    %% Forall_R
    \begin{array}{c}
    \infer{\vdash^{d}\Gamma,\forall x A\gives\sig}{\vdash^{\addFV{d}{\EVar{x}}}\Gamma,A\left[x:=\EVar{x}\right]\gives\sig}\\
    \mbox{\scriptsize where $\EVar{x}$ is a fresh eigenvariable}\\
    \end{array}
    \\
  \end{array}}
  \]
  \caption{The constraint-producing sequent calculus \DI}
  \label{sys2}
\end{figure}

In terms of process, a sequent $\vdash^{d}\Gamma\gives \sigma$
displays the inputs $\Gamma, d$ and the output $\sigma$ of proof
search, which starts building a proof tree, in system \DI, from the
root. The sequent at the root would typically be of the form
$\vdash^{\dinit}\Gamma\gives\sigma$, with $\sigma\in\Psi_{\dinit}$ to
be produced as output. The constraints are produced at the leaves, and
propagated back down towards the root.
\begin{example}\label{ex:cppred}
  In
  Examples~\ref{ex:pure1st-order},~\ref{ex:eq},~\ref{ex:ground},
  the constraint-producing predicate $\BBc{\mathcal A}\sig d$ holds
  if, respectively, $\sigma$ is the most general unifier of two dual
  literals in $\mathcal A$, $\sigma$ is an output of rigid
  E-unification on $\mathcal A$, $\sigma$ is a ground substitution for
  which $\sigma(\mathcal A)$ is $\mathcal T$-inconsistent.
  In Example~\ref{ex:qe}, $\BBc{\mathcal A}\sig d$ holds if the
  quantifier-free formula $\sigma$ (of appropriate domain) implies
  $\mathcal A$ (as a disjunction). 
  For our specific example, which
  also involves uninterpreted predicate symbols, proof search in
  system \DI\ builds a tree
  \[\small
  \infers{
    \vdash^{d_0}(\exists xy(p(x,y)\wedge l(x,y)))\vee
    (\exists x'y'(\non p(x',y')\wedge l'(x',y')))\gives
    \sig_{\downarrow\downarrow\downarrow\downarrow}
  }
  {
    \infers{\ldots}
    {
      \infers{
        \vdash^{d}(p(\XVar X,\XVar Y)\wedge l(\XVar X,\XVar Y)),
        (\non p(\XVar {X'},\XVar {Y'})\wedge l'(\XVar {X'},\XVar {Y'}))\gives \sig
      }
      {
        \infers{\ldots}
        {
          \begin{array}{cc}
            \vdash^{d} p(\XVar X,\XVar Y),\non p(\XVar {X'},\XVar {Y'})\gives
            \sig_1
            &\quad
            \vdash^{d} l(\XVar X,\XVar Y),\non p(\XVar {X'},\XVar {Y'})\gives
            \sig_2\\
            \vdash^{d} p(\XVar X,\XVar Y),l'(\XVar {X'},\XVar {Y'})\gives
            \sig_3
            &\quad
            \vdash^{d} l(\XVar X,\XVar Y),l'(\XVar {X'},\XVar {Y'})\gives
            \sig_4
          \end{array}
        }
      }
    }
  }
  \] 
  where $d \eqdef \XVar X; \XVar Y; \XVar X'; \XVar Y'$, the
  background reasoner produces $\sig_1\eqdef\{\XVar X = \XVar X';
  \XVar Y = \XVar Y'\}$, $\sig_2\eqdef\{3\XVar X \leq 2 \XVar Y \leq
  3\XVar X + 1\}$, $\sig_3\eqdef\{99 \leq 3 \XVar Y' +2 \XVar X' \leq
  101\}$, and $\sig_4\eqdef\sig_2$ ($\sig_4\eqdef\sig_3$ also works);
  then $\sig \eqdef (\sig_1 \meet \sig_2) \meet (\sig_3 \meet \sig_4)$
  and finally $\sig_{\downarrow\downarrow\downarrow\downarrow}$,
  obtained by quantifier elimination from $\sig$, is the trivially
  true formula.
% rooted with the sequent
% \[\vdash^{\emptyset}(\exists xy(p(x,y)\wedge l(x,y)))\vee
%   (\exists x'y'(\non p(x',y')\wedge l'(x',y')))\gives
%   \sig_{\downarrow\downarrow\downarrow\downarrow}\]
% for some constraint $\sig$: after applying the rules
% for disjunction and existential quantifier, a proof tree is built for
%  \[\vdash^{d}(p(x,y)\wedge l(x,y))\vee
%   (\non p(x',y')\wedge l'(x',y'))\gives \sig\]
% with $d \eqdef \XVar X; \XVar Y; \XVar X'; \XVar Y'$. Now $\sig$ is of the
% form $(\sig_1 \meet \sig_2) \meet (\sig_3 \meet \sig_4)$ because from
% this point, the proof tree opens branches whose leaves are
% respectively labelled with sequents:
%   \[\begin{array}{cc}
%     \vdash^{d} p(\XVar X,\XVar Y),\non p(\XVar {X'},\XVar {Y'})\gives
%   \sig_1
%     &\quad
%     \vdash^{d} l(\XVar X,\XVar Y),\non p(\XVar {X'},\XVar {Y'})\gives
%   \sig_2\\
%     \vdash^{d} p(\XVar X,\XVar Y),l'(\XVar {X'},\XVar {Y'})\gives
%   \sig_3
%     &\quad
%     \vdash^{d} l(\XVar X,\XVar Y),l'(\XVar {X'},\XVar {Y'})\gives
%   \sig_4
%   \end{array}
%   \]
% The constraint-producing predicate called at these leaves instanciates
% the constraints: for instance
% $\sig_1$ is set at $\{\XVar X = \XVar X'; \XVar Y = \XVar Y'\}$ and
% $\sig_4$ at $\{3\XVar X \leq 2 \XVar Y \leq 3\XVar X + 1;
% 99 \leq 3 \XVar Y' +2 \XVar X' \leq 101\}$.
\end{example}
System \DI\ is very close to Rümmer's PresPred${}^C_S$
System~\cite{princess08}, but using abstract constraints instead of linear
arithmetic constraints. Using \DI\ with the constraint structure of
Example~\ref{ex:constraintstruct}.3 implements
Rümmer's suggestion~\cite{princess08} to eliminate quantifiers along the
propagation of constraints down to the root.

\longversion{
  Given that the initial domain corresponds to the absence of
  meta-variables, a typical example for $\Psi_{\dinit}$ is the set of
  booleans (a constraint of $\Psi_{\dinit}$ simply being either the
  satisfiable constraint or the unsatisfiable constraint).
}

\subsection{Instantiations and Compatibility with Constraints}

Notice that, in system \DI, no instantiation for meta-variables is
actually ever produced. Instantiations would only come up when
reconstructing, from an \DI\ proof, a proof in the original calculus
\LKone. So as to relate constraints to actual instantiations, we
formalise what it means for an instantiation to satisfy, or be
compatible with, a constraint of domain $d$. Such an instantiation
should provide, for each meta-variable, a term that at least respects
the eigenvariable dependencies specified in $d$, as formalised in
Definition~\ref{def:instsubst}. Beyond this, what it means for an
instantiation to be compatible with a constraint is specific to the
theory and we simply identify in Definition~\ref{def:compatrel} some
minimal axioms.  We list these axioms, along with the rest of this
paper's axiomatisation, in Fig.~\ref{fig:summary} on
page~\pageref{fig:summary}.

\begin{definition}[Instantiation]\label{def:instsubst}

  The set of \Index{instantiations of domain $d$}, denoted $\Inst{d}$, is the
  set of mappings from meta-variables to ground terms defined by
  induction on $d$ as follows:
  \[
      \Inst{\dinit} = \emptyInst \qquad
      \Inst{\addFV{d}{\EVar{x}}} = \Inst{d} \qquad
      \Inst{\addFV{d}{\XVar{X}}} = \left\{\addInst{\XVar{X}}{t}{\rho} \mid t \in T_d, \rho \in \Inst{d}\right\}
  \]

  For a term $t$ (\resp a formula $A$, a context $\Gamma$) of domain
  $d$ and an instantiation $\rho\in \Inst{d}$, we denote by $\rho(t)$
  (\resp $\rho(A)$, $\rho(\Gamma)$) the result of substituting in $t$
  (\resp $A$, $\Gamma$) each meta-variable $\XVar{X}$ in $d$
  by its image through $\rho$.
\end{definition}

\begin{definition}[Compatibility relation]\label{def:compatrel}
  A \Index{compatibility relation} is a (family of) relation(s)
  between instantiations $\rho\in \Inst{d}$ and constraints
  $\sigma\in\Psi_d$ for each domain $d$, denoted $\rho\compat\sig$,
  that satisfies Axiom \AXproj\ of Fig.~\ref{fig:summary}.

  If the constraint structure is a meet constraint structure, we say
  that the compatibility relation \Index{distributes} over $\meet$ if
  it satisfies Axiom \AXmeet\ of Fig.~\ref{fig:summary}.
\end{definition}

%% REMOVED FOLLOWING AXIOM, as it is consequence of C1 right-to-left (which itself was a consequence of this axiom)
%% \forall\sig,\sig',\ \sig\stronger\sig'\Rightarrow \left\{\rho\mid\rho\compat\sig\right\}\subseteq
%% \left\{\rho\mid\rho\compat\sig'\right\}\\

Another ingredient we need to relate the two sequent calculi is a mechanism for
producing instantiations. We formalise a \emph{witness builder} which maps
every constraint of $\Psi_{\addFV{d}{\XVar{X}}}$ to a function, which outputs
an ``appropriate'' instantiation for $\XVar{X}$ when given as input an
instantiation of domain $d$:

\begin{definition}[Witness]
  A \Index{witness builder} for a compatibility relation $\compat$ is
  a (family of) function(s) that maps every
  $\sigma\in\Psi_{\addFV{d}{\XVar{X}}}$ to $\bind \sigma\in
  \Inst{d}\rightarrow T_d$, for every domain $d$ and every
  meta-variable $\XVar{X}$, and that satisfies Axiom \AXwitness\ of
  Fig.~\ref{fig:summary}.
  % \[
  %% \begin{array}{ll}
  %% \textsf{B2}&\forall c\in\Psi_{q},\ \forall\rho\compat\left(\sig'\meet c\right),\ \left(f\left(\rho\right)\cons
  %% \rho\right)\compat\left(\sig\meet\tilde{c}\right)\qquad \mbox{[SGL: $\meet c$ unnecessary with \textsf{C4}]}\\
  %% \textsf{B3}&\sigma\stronger\inject{{\sigma'}}\qquad \mbox{[SGL: stronger version of \textsf{B1}, using \textsf{C4}, unnecessary?]}
  %% \end{array}
  % \]
\end{definition}

\begin{example}
  For the constraint structure of Example~\ref{ex:constraintstruct}.1,
  we can define: $\rho\compat\sig$ if $\rho$ is a
  (ground) instance of substitution $\sigma$. Given
  $\sig\in\Psi_{\addFV{d}{\XVar{X}}}$ and $\rho\compat\project\sig$,
  we have $\rho=\rho'\circ\project\sig$, and we can take $\bind \sigma
  (\rho)$ to be any instance of $\rho'(\sigma(\XVar{X}))$ in $\Inst{d}$.

  In the particular case of Example~\ref{ex:constraintstruct}.2,
  $\rho\compat\sig$ if $\rho$ coincides with $\sigma$ (on every
  meta-variable not mapped to itself by $\sigma$). To define $\bind
  \sigma (\rho)$, note that $\rho'(\sigma(\XVar{X}))$ is either ground
  or it is $\XVar{X}$ itself (in which case any term in $\Inst{d}$
  works as $\bind \sigma (\rho)$).

  For the constraint structure of Example~\ref{ex:constraintstruct}.3,
  we can take: $\rho \compat
  F$ if the ground formula $\rho(F)$ is valid in the theory. From a
  formula $F \in\Psi_{\addFV{d}{\XVar{X}}}$ and an instantiation
  $\rho$, the term $\bind F (\rho)$ should represent an existential
  witness for the formula $\rho(F)$, which features $\XVar{X}$ as the
  only meta-variable. In the general case, we might need to resort to
  a Hilbert-style choice operator to construct the witness:
  $\epsilon(\exists x((\addInst{\XVar{X}}x{\rho})(F))$. For instance
  in the case of linear arithmetic, $\rho(F)$ corresponds to a
  disjunction of systems of linear constraints, involving $\XVar{X}$
  and the eigenvariables $\vect{\EVar y}$ of $d$. Expressing how
  $\XVar{X}$ functionally depends on $\vect{\EVar y}$ to build a
  solution of one of the systems, may require extending the syntax of
  terms. But note that proof search in \DI\ does not require
  implementing a witness builder.
%  In the case of Example~\ref{ex:qe} for a theory with quantifier
%  elimination, we need that the process turning a formula $\exists x
%  F$ (with $F$ being quantifier-free) into an equivalent
%  quantifier-free formula $F'$ effectively produces a witness term $t$
%  such that $F[x:=t]$ is valid if $F'$ is valid. In this case for
%  $F\in\Psi_{n\cons l}$, $\bind F (\rho)$ is $\rho(t)$ where $t$ is
%  the witness term for the quantifier elimination transforming
%  $\exists x(F[\XVar{\abs l + 1}:=x])$ into an equivalent quantifier-free formula.

  A meet constraint structure can also be defined by taking
  constraints to be (theory-specific kinds of) sets of instantiations:
  the compatibility relation $\compat$ is just set membership, set
  intersection provides a meet operator and the projection of a
  constraint is obtained by removing the appropriate entry in every
  instantiation belonging to the constraint. Witness building would
  still be theory-specific.
\end{example}

To relate Systems \DI\ and \LKone, we relate constraint-producing
predicates to ground validity ones: intuitively, the instantiations
that turn a set $\mathcal A$ of literals of domain $d$ into a valid
set of (ground) literals should coincide with the instantiations that
are compatible with some constraint produced for $\mathcal A$ (a
similar condition appears in Theorem~55 of~\cite{beckert99handbook}
with $\mathcal T$-refuters instead of constraints):

\begin{definition}[Relating predicates]\label{def:blackbox-rel}

  For a compatibility relation $\compat$, we say that a
  constraint-producing predicate $(\models^d)_d$ \Index{relates to} a
  ground validity predicate $\models$ if they satify
  Axiom \AXpg\ of Fig.~\ref{fig:summary}.
\end{definition}

A constraint-producing predicate may allow several constraints to
close a given leaf (finitely many for Example~\ref{ex:pure1st-order},
possibly infinitely many for Examples~\ref{ex:eq} and~\ref{ex:ground},
just one for Example~\ref{ex:qe}). So in general our foreground
reasoner expects a \emph{stream} of constraints to be produced at
a leaf, corresponding to the (possibly infinite) union in axiom
\AXpg: each one of them is sufficient to close the branch.  The first
one is tried, and if it later proves unsuitable, the next one
in the stream can be tried, following Giese's suggestion~\cite{Giese00}
of using streams of instantiations.

\subsection{Soundness and Completeness}

\shortversion{ System \DI\ can be proved equivalent to System \LKone,
  from the axioms in the top half of
  Fig.~\ref{fig:summary}~\cite{rouhling:hal-01107944}.  To state this
  equivalence, assume that we have a compatibility relation that
  distributes over $\meet$, equipped with a witness builder, plus a
  constraint-producing predicate $(\models^d)_d$ related to a ground
  validity predicate $\models$. }

\longversion{We now prove the equivalence between System \DI\ and System \LKone,
  from the axioms in the top half of Fig.~\ref{fig:summary}.
  We hence assume that we have a compatibility
  relation that distributes over $\meet$, equipped with a witness
  builder, plus a constraint-producing predicate $(\models^d)_d$
  related to a ground validity predicate $\models$.}

\begin{toappendix}
  \begin{theorem}[Soundness and completeness of \DI]\strut\label{th:BB1sound}
    \label{th:BB1complete}

    For all contexts $\Gamma$ of domain $d$:

    If $\vdash^{d}\Gamma\gives \sigma$ is derivable in \DI, then
    for all $\rho\compat\sigma$, $\vdash\rho(\Gamma)$ is derivable
    in \LKone.

    For all $\rho\in \Inst{d}$, if $\vdash\rho(\Gamma)$ is derivable in
    \LKone, then there exists $\sigma\in\Psi_d$ such that
    $\vdash^{d}\Gamma\gives \sigma$ is derivable in \DI\ and
    $\rho\compat\sigma$.
  \end{theorem}
\end{toappendix}

\longversion{
  \begin{toappendix}[\begin{proof} See the proofs in Appendix~\thisappendix.\end{proof}]
    \begin{proof}
      We first prove the soundess of {\DI} by induction on the derivation
      of $\vdash^{d}\Gamma\gives\sig$:

      \begin{itemize}
      \item[Theory]
        \[\infer[\BBc{\atmCtxt\Gamma}\sig d]{\vdash^{d}\Gamma\gives\sig}{}\]
        Let $\rho\compat\sig$. By $\AXpg$ (right-to-left inclusion), we have $\rho\in\left\{\rho\mid\;  \BB{\rho\left(\atmCtxt\Gamma\right)}\right\}$.
      \item ghj%[\strut $\exists$-intro]\strut
        \[
        \infer
        {
          \vdash^{d}\Gamma,\exists x A\gives\project\sig
        }
        {
          \vdash^{\addFV{d}{\XVar{X}}}\Gamma,A\left[x:=\XVar{X}\right], \exists x A\gives\sig}
        \]
        Let $\rho\compat\project\sigma$. By Axiom \AXwitness, we have $(\addInst{\XVar{X}}{\bind\sigma(\rho)}{\rho})\compat\sigma$ and applying the induction hypothesis we can construct
        \[\infer{\vdash\rho\left(\Gamma\right),\rho(\exists x A)}
        {\vdash\rho\left(\Gamma\right),(\rho(A))[x:=\bind \sigma(\rho)], \rho(\exists x A)\gives\sig}
        \]

      \item[$\wedge$-intro]
        \[\infer{\vdash^{d}\Gamma,A\wedge B\gives\sig\meet\sig'}
        {\vdash^{d}\Gamma,A\gives\sig \qquad \vdash^{d}\Gamma,B\gives\sig'}
        \]
        Let $\rho\compat\sig\meet\sig'$.
        By Axiom \AXmeet\ (right-to-left), we have $\rho\compat\sig$ and $\rho\compat\sig'$
        and we can conclude by applying the induction hypothesis.
      \item[$\vee$-intro]
        \[\infer{\vdash^{d}\Gamma,A\vee B\gives\sig}{\vdash^{d}\Gamma,A,B\gives\sig}\]
        We conclude immediately with the induction hypothesis.
      \item[$\forall$-intro]
        \[\infer{\vdash^{d}\Gamma,\forall x A\gives\sig}{\vdash^{\addFV{d}{\EVar{x}}}\Gamma,A\left[x:=\EVar{x}\right]\gives\sig}\]
        We conclude immediately with the induction hypothesis.
      \end{itemize}

      We now prove the completeness result.

      Consider a domain $d$, an instantiation $\rho\in \Inst{d}$ and a
      context $\Gamma$ of domain $d$. The proof goes by induction on the
      derivation of $\vdash\rho(\Gamma)$ in the original calculus:
      \begin{itemize}
      \item[Theory]
        \[\infer[\BB{\atmCtxt\Gamma}]{\vdash\rho(\Gamma)}{}\]
        Since the constraint-producing predicate relates to the ground
        validity predicate, by Definition~\ref{def:blackbox-rel} there
        exists a constraint $\sigma \in \Psi_d$ such that:
        \[
        \BBc{\atmCtxt\Gamma}\sig d\textrm{ and }
        \rho\compat\sig
        \]
        The proof follows by applying the theory rule of the {\DI} system.

      \item[$\exists$-intro]
        Note that instantiations do not feature meta-variables,
        hence for any context $\Gamma$ and any formula $A$, $\rho(\Gamma,
        \exists x A)$ is syntactically equal to $(\rho(\Gamma),\exists x\,
        \rho(A))$. In the present case of the induction, the derivation
        hence ends with a rule of the form:
        \[
        \infer{\vdash\rho(\Gamma),\exists x\,\rho(A)}{\vdash\rho(\Gamma),
          \rho(A)\left[x:=t\right],\exists x \rho(A)}
        \]

        Introduce the instantiation $\rho' := \addInst{\XVar{X}}{t}{\rho}
        \in \Inst{\addFV{d}{\XVar{X}}}$. The formula
        $\rho(A)\left[x:=t\right]$ is syntactically equal to
        $\rho'(A\left[x:=\XVar{X} \right])$. The premiss of the former
        rule can hence be written as:
        \[
        \vdash\rho'(\Gamma),\rho'(A\left[x:=\XVar{X}\right]),\exists x \rho'(A)
        \]
        which is in turn of the form:
        \[
        \vdash\rho'(\Gamma,A\left[x:=\XVar{X}\right],\exists x A)
        \]

        By induction hypothesis, there exists $\sigma'$ such that:
        \[
        \vdash^{\addFV{d}{\XVar{X}}}\Gamma,A\left[x:=\XVar{X}\right],\exists x (A) \into
        \sigma' \textrm{ and }\rho'  \compat \sigma'
        \]

        Therefore $\vdash^d\Gamma, \exists x A \into \sigma$ with
        $\sigma = \project \sigma'$ and $\rho \compat \sigma$ follows from Axiom
        \AXproj.
      \item[$\wedge$-intro]
        \[\infer{\vdash\rho(\Gamma),\rho(A)\wedge \rho(B)}{\vdash\rho(\Gamma),\rho(A) \qquad \vdash\rho(\Gamma),\rho(B)}\]
        By induction hypothesis, there exist $\sigma_1,\sigma_2 \in
        \Psi_d$ such that:
        \[\rho\compat \sigma_1, \quad \rho\compat
        \sigma_2, \quad \vdash^d\Gamma,A \into \sigma_1, \quad
        \vdash^d\Gamma,B \into \sigma_2\]
        Hence $\vdash^d\Gamma, A\wedge B \into \sigma_1 \meet
        \sigma_2$ and $\rho\compat \sigma_1 \meet \sigma_2$ follows from Axiom
        \AXmeet\ (left-to-right).
      \item[$\vee$-intro]

        We conclude immediately by induction hypothesis.
      \item[$\forall$-intro]

        We conclude immediately by induction hypothesis since instantiations do not
        affect eigenvariables.
      \end{itemize}

    \end{proof}
  \end{toappendix}
}

We will usually start proof search with the domain $\dinit$, so as to
build a proof tree whose root is of the form
$\vdash^{\dinit}\Gamma\into\sigma$ for some constraint
$\sigma\in\Psi_{\dinit}$. Since the only instantiation in
$\Inst{\dinit}$ is $\emptyInst$, and since
$\emptyInst(\Gamma)=\Gamma$, soundness and completeness for domain
$\dinit$ can be rewritten as follows:

\begin{corollary}[Soundness and completeness for the initial domain]\label{cor:SDIsoundcomplempty}

  There exists $\sigma\in\Psi_{\dinit}$ such that
  $\vdash^{\dinit}\Gamma\gives \sigma$ is derivable in
  \DI\ and $\emptyInst\compat\sigma$,\\
  if and only if $\vdash\Gamma$ is derivable in \LKone.
\end{corollary}

\longversion{
  \subsection{Canonical instantiation}

  Interestingly enough, a witness builder elects a canonical
  instantiation among all of those that are compatible with a
  given constraint, by iterating projections down to the empty domain as
  follows:

  \begin{definition}[Folding a constraint]
    Given a witness builder, the \Index{fold} of a constraint
    $\sigma\in\Psi_d$ is an instantiation of $\Inst{d}$ defined by induction
    on $d$:
    \[
    \begin{array}{lll}
      \fold{(\sig)} &\eqdef \emptyset & \mbox{if $\sigma\in\Psi_{\dinit}$}\\
      % \fold{(\sig)} &\eqdef \fold{(\project\sig)} &
      % \mbox{if $\sigma\in\Psi_{\addFV{d}{\EVar{x}}}$}\\
      \fold{(\sig)} &\eqdef
      \addInst{\XVar{X}}{\bind\sig(\fold{(\project\sig)})}{\fold{(\project\sig)}} &
      \mbox{if $\sigma\in\Psi_{\addFV{d}{\XVar{X}}}$}
    \end{array}
    \]
  \end{definition}

  \begin{remark}
    The main information about a constraint of empty domain
    $\Psi_{\dinit}$ is whether $\emptyset$, the unique instantiation in
    $\Inst{\dinit}$, is compatible with it or not.

    For every $d$ and every $\sigma\in\Psi_d$, there exists
    $\rho$ such that $\rho\compat\sig$ \iff\ $\fold{(\sig)}\compat \sig$
    (the left-to-right direction is proved by induction on $d$ using
    axioms \AXproj\ and \AXwitness). When this is the case, we
    say that $\sigma$ is \Index{satisfiable}.
  \end{remark}

  This can be directly applied to the canonical instantiation given by
  $\fold{(\sig)}$:

  \begin{corollary}[Soundness with canonical instantiation]

    If $\vdash^{d}\Gamma\gives \sigma$ is derivable in \DI, with
    $\sigma$ being satisfiable, then $\vdash\fold{(\sig)}(\Gamma)$
    is derivable in \LKone.
  \end{corollary}

}
\section{Sequentialising}
\label{sec:sequentialising}

The soundness and completeness properties of System \DI\ rely on
constraints that are satisfiable. A proof-search process based on it
should therefore not proceed any further with a constraint that has
become unsatisfiable. Since the meet of two satisfiable constraints
may be unsatisfiable, % exploring the two branches created by 
branching on conjunctions may take advantage of a sequential
treatment: a constraint produced to close one branch may direct the
exploration of the other branch, which may be more efficient than
waiting until both branches have independently produced constraints
and only then checking that their meet is satisfiable. This
section develops a variant of System \DI\ to support this
sequentialisation of branches, much closer than System \DI\ to the
free variable tableaux with theory reasoning~\cite{beckert99handbook}.

In the rest of this section  $(\Psi_{d})_d$ is a fixed lift
constraint structure.

\subsection{Definition of the Proof System}

Thus, the proof rules enrich a sequent with \emph{two} constraints:
the input one and the output one, the latter being ``stronger'' than
the former, in a sense that we will make precise when we relate the
different systems. At the leaves, a new predicate $(\BBri{d})_d$ is
used that now takes an extra argument: the input constraint.

\begin{definition}[{\SDI} sequent calculus]

  A \emph{constraint-refining predicate} is a family of relations
  $(\BBri{d})_d$, indexed by domains $d$, relating sets $\mathcal A$
  of literals of domain $d$ with pairs of constraints $\sigma$ and
  $\sigma'$ in $\Psi_d$; when it holds, we write
  $\BBr{\sigma}{\mathcal A}{\sigma'}{d}$.

  Given such a predicate $(\BBri{d})_d$, the
  \Index{constraint-refining sequent calculus}, denoted {\SDI},
  manipulates sequents of the form
  $\sig\into\vdash^{d}\Gamma\gives\sig'$, where $\Gamma$ is a context
  and $\sigma$ and $\sigma'$ are constraints, all of domain $d$.  Its
  rules are presented in Fig.~\ref{sys3}.
\end{definition}

\begin{figure}[h]
\[
\boxed{\begin{array}{c}
%% Axiom
\infer[\mbox{$\BBr{\sig}{\,\atmCtxt\Gamma}{\sig'}{d}$}]{\sig\into\vdash^{d}\Gamma\gives\sig'}{}
\qquad
%% Or_R
\infer{\sig\into\vdash^{d}\Gamma,A\vee
B\gives\sig'}{\sig\into\vdash^{d}\Gamma,A,B\gives\sig'}
\\\\

%% And_R
\infer[i\in\{0,1\}]{\sig\into\vdash^{d}\Gamma,A_{0}\wedge
A_{1}\gives\sig'}{\sig\into\vdash^{d}\Gamma,A_{i}\gives\sig'' \qquad
\sig''\into\vdash^{d}\Gamma,A_{1-i}\gives\sig'}
\\\\

%% Exists_R
\begin{array}{c}

\infer{\sig\into\vdash^{d}\Gamma,\exists x A\gives\project
{\sig'}}{\inject{\sig}\into\vdash^{\addFV{d}{\XVar{X}}}\Gamma,A\left[x:=\XVar{X}\right],\exists
x A\gives\sig'}\\
\mbox{\scriptsize where $\XVar{X}$ is a fresh meta-variable}\\
\end{array}
\qquad

%% Forall_R
\begin{array}{c}
\infer{\sig\into\vdash^{d}\Gamma,\forall x
A\gives\sig'}{\sig\into\vdash^{\addFV{d}{\EVar{x}}}\Gamma,A\left[x:=\EVar{x}\right]\gives\sig'}
\\
\mbox{\scriptsize where $\EVar{x}$ is a fresh eigenvariable}\\
\end{array}\\
\end{array}}
\]
\caption{The sequent calculus with sequential delayed instantiation {\SDI}}
\label{sys3}
\end{figure}

The branching rule introducing conjunctions allows an arbitrary
sequentialisation of the branches when building a proof tree, proving
$A_0$ first if $i=0$, or proving $A_1$ first if $i=1$.

\begin{example}
  In Examples~\ref{ex:pure1st-order},~\ref{ex:eq},~\ref{ex:ground},
  constraints are simply substitutions, and the constraint-refining
  predicate $\BBr{\sig}{\,\mathcal A}{\sig'}{d}$ is taken to hold if
  the constraint-producing predicate $\BBc{\sig(\mathcal A)}{\sig'} d$
  (as given in Example~\ref{ex:cppred}) holds. Here we recover the
  standard behaviour of free variable tableaux (with or without
  theory~\cite{beckert99handbook}) where the substitutions used to
  close branches are applied to the literals on the remaining
  branches. Of course in both cases, an implementation may apply the
  substitution lazily. In Example~\ref{ex:qe}, the constraint-refining
  predicate $\BBr{\sig}{\,\mathcal A}{\sig'}{d}$ is taken to hold if
  $\BBc{(\sig\wedge\mathcal A)}{\sig'} d$ holds.  Proof search in
  \SDI\ builds, for our specific example and a trivially true
  constraint $\sig_0$, the proof-tree
  \[\small
  \infers{
    \sig_0\into\vdash^{d_0}(\exists xy(p(x,y)\wedge l(x,y)))\vee
    (\exists x'y'(\non p(x',y')\wedge l'(x',y')))\gives
    \sig'_{\downarrow\downarrow\downarrow\downarrow}
  }
  {
    \infers{\ldots}
    {
      \infers{
        \sig_0\into\vdash^{d}(p(\XVar X,\XVar Y)\wedge l(\XVar X,\XVar Y)),
        (\non p(\XVar {X'},\XVar {Y'})\wedge l'(\XVar {X'},\XVar {Y'}))\gives \sig'
      }
      {
        \infers{\ldots}
        {
          \begin{array}{cc}
            \sig_0\into\vdash^{d} p(\XVar X,\XVar Y),\non p(\XVar {X'},\XVar {Y'})\gives
            \sig_1'
            &\quad
            \sig_1'\into\vdash^{d} l(\XVar X,\XVar Y),\non p(\XVar {X'},\XVar {Y'})\gives
            \sig_2'\\
            \sig_2'\into\vdash^{d} p(\XVar X,\XVar Y),l'(\XVar {X'},\XVar {Y'})\gives
            \sig_3'
            &\quad
            \sig_3'\into\vdash^{d} l(\XVar X,\XVar Y),l'(\XVar {X'},\XVar {Y'})\gives
            \sig'
          \end{array}
        }
      }
    }
  }
  \] 
  similar to that of Example~\ref{ex:cppred}, where $\sigma_1'\eqdef\sigma_1$, $\sigma_2'\eqdef\sigma_1'\wedge\sigma_2$, 
$\sigma_3'\eqdef\sigma_2'\wedge\sigma_3$ and $\sigma'\eqdef\sigma_3'$, projected by quantifier elimination to the trivially true formula $\sig'_{\downarrow\downarrow\downarrow\downarrow}$.
\end{example}

\subsection{Soundness and Completeness}

We now relate system {\SDI} to system {\DI}. For this we need some axioms
about the notions used in each of the two systems. These are distinct from the
axioms that we used to relate system {\DI} to {\LKone}, since
we are not (yet) trying to relate system {\SDI} to {\LKone}.
In the next section however, we will combine the two steps.

\begin{definition}[Decency]
  When $\stronger$ (\resp $\meet$, $P$) is a family of pre-orders\linebreak
  (\resp binary operators, predicates) over each $\Psi_d$, we say that
  $(\stronger,\meet,P)$ is \Index{decent} if the following axioms
  hold:
  \[
    \begin{array}{c@{\qquad\qquad}c}
      %% Axiom D1
      \multicolumn{2}{l}{
      \textsf{D1}\quad
      \forall \sigma,\sigma'\in\Psi_d,\;\sig\wedge\sig'\mbox{ is a
      greatest lower bound of $\sig$ and $\sig'$ for $\stronger$}}\\
      %% Axiom D2
      \multicolumn{2}{l}{
      \textsf{D2}\quad \forall \sigma\in\Psi_d\,\forall
      \sigma',\sigma''\in\Psi_{\addFV{d}{\XVar{X}}},\;
      \sigma''\obseq\inject\sigma\meet\sigma'\ \Rightarrow\ \project{\sigma''}\
      \obseq\ \sigma\meet\project{\sigma'}}\\
      %% Axiom P1
      \textsf{P1}\quad \forall\sigma\in\Psi_{\addFV{d}{\XVar{X}}},\;P(\sig)\Leftrightarrow P(\project\sig) &
      %% Axiom P2
      \textsf{P2}\quad \forall\sigma,\sigma'\in\Psi_d,\;\left\{
      \begin{array}{l}
        P(\sig)\\
        \sig\stronger\sig'\\
      \end{array}\right.\Rightarrow P(\sig')\\
      %% \textsf{P3}& \forall\sigma\in\Psi_l,\forall\sigma'\in\Psi_{n\cons l},&
      %% P(\sigma\meet\project{\sigma'})\ \Rightarrow\
      %% P(\inject\sigma\meet\sigma')
    \end{array}
  \]
  where $\obseq$ denotes the equivalence relation generated by $\stronger$.
\end{definition}

Notice that this makes $(\sfrac{\Psi_d}{\obseq,\meet})$ a
meet-semilattice that could equally be defined by the associativity,
commutativity, and idempotency of $\meet$.

\begin{definition}[Relating constraint-producing/refining predicates]

  Given a family of binary operators $\meet$ and a family of predicates $P$, we
  say that a constraint-refining predicate $(\BBri{d})_d$ \Index{relates to} a
  constraint-producing predicate $(\BBci{d})_d$
  if, for all domains $d$, all sets $\mathcal A$ of literals of domain $d$
  and all $\sigma\in\Psi_d$,
  \[
  \begin{array}{l@{\quad}lll}
    %% Axiom A1
    \textsf{A1}
    &
    \forall \sigma'\in\Psi_d,\quad \BBr{\sig}{\mathcal A}{\sig'}{d}
    &\Rightarrow\ \exists \sig''\in\Psi_d,
    \left\{
      \begin{array}{l}
      \sigma'\obseq\sig\meet\sig''\\
      P(\sig\meet\sig'')\\
      \BBc{\mathcal A}{\sig''}{d}
    \end{array}
    \right.\\
    %% Axiom A2
    \textsf{A2}
    &
    \forall \sigma'\in\Psi_d,\quad
    \left\{
      \begin{array}{l}
      P(\sig\meet\sig')\\
      \BBc{\mathcal A}{\sig'}{d}
    \end{array}
    \right.
    &\Rightarrow\ \exists \sigma''\in\Psi_d,
    \left\{
      \begin{array}{l}
      \sigma''\obseq\sig\meet\sig'\\
      \BBr{\sig}{\mathcal A}{\sig''}{d}
    \end{array}
    \right.
  \end{array}
  \]
\end{definition}

In the rest of this sub-section, we assume that we have a decent
triple $(\stronger, \meet, P)$, and a constraint-refining predicate
$(\BBri{d})_d$ that relates to a contraint-producing predicate
$(\BBci{d})_d$.  In this paper we only use two predicates $P$,
allowing us to develop two variants of each theorem, with a compact
presentation: $P(\sig)$ is always ``true'', and $P(\sig)$ is ``$\sig$
is satisfiable'', both of which satisfy \textsf{P1} and \textsf{P2}.

\shortversion{
  System \SDI\ can then be proved sound with respect to System \DI~\cite{rouhling:hal-01107944}:
}

\begin{toappendix}[\label{cor:soundnesscanonical}]
  \subsection{Proof of Soundness for \SDI}
\end{toappendix}

\begin{toappendix}
  \begin{theorem}[Soundness of {\SDI}]\label{th:BB2sound}

  If $\sigma\into\vdash^{d}\Gamma\gives \sigma'$ is derivable in {\SDI},
  then there exists $\sigma''\in\Psi_d$ such that
  $\sigma'\obseq\sigma\meet\sigma''$, $P(\sigma\wedge\sigma'')$ and
  $\vdash^{d}\Gamma\gives \sigma''$ is derivable in {\DI}.
  \end{theorem}
\end{toappendix}

\longversion{
  \begin{toappendix}[\begin{proof} See the proof in Appendix~\thisappendix.\end{proof}]
    \begin{proof}
      By induction on the derivation of $\sigma\into\vdash^{d}\Gamma\gives \sigma'$:
      \begin{itemize}

      \item[Theory]
        \[
        \infer[\mbox{$\BBr{\sigma}{\atmCtxt\Gamma}{\sigma'}{d}$}]{\sigma\into
          \vdash^{d}\Gamma\gives\sigma'}{}
        \]
        By $\AXrp$, there exists $\sigma''\in\Psi_{d}$ such that
        $\sigma'\obseq\sigma\meet\sigma''$, $P\left(\sigma\meet\sigma''\right)$
        and $\BBc{\atmCtxt\Gamma}{\sigma''}{d}$. We can then immediately
        conclude.

      \item[$\exists$-intro]
        \[
        \infer{\sigma\into\vdash^{d}\Gamma,\exists x
          A\gives\project{\sigma'}}{\inject{\sigma}\into\vdash^{\addFV{d}{\XVar{X}}}
          \Gamma,A\left[x:=\XVar{X}\right],\exists x A\gives\sigma'}
        \]
        By the induction hypothesis, there exists
        $\sigma''\in\Psi_{\addFV{d}{\XVar{X}}}$ such that
        $$\sigma'\obseq\inject{\sigma}\meet\sigma'',\
        P\left(\inject{\sigma}\meet\sigma''\right)\text{ and }
        \vdash^{\addFV{d}{\XVar{X}}}\Gamma,A\left[x:=\XVar{X}\right],\exists
        xA\gives\sigma''$$ is derivable in {\DI}. Thus,
        $$\vdash^{d}\Gamma,\exists xA\gives\project{\sigma''}$$ is derivable
        in {\DI}. Let us then show that $\project{\sigma''}$ satisfies each
        constraint:

        Since $\sigma''\in\Psi_{\addFV{d}{\XVar{X}}}$, we have
        $\project{\sigma''}\in\Psi_{d}$. By $\textsf{D2}$, knowing
        $\sigma'\obseq\inject{\sigma}\meet\sigma''$, we get
        $\project{\sigma'}\obseq\sigma\meet\project{\sigma''}$. Since
        $\inject{\sigma}\meet\sigma''\stronger\sigma'$ and
        $P\left(\inject{\sigma}\meet\sigma''\right)$, by $\textsf{P2}$ we get
        $P\left(\sigma'\right)$. Then, by $\textsf{P1}$ (left-to-right),
        $P\left(\project{\sigma'}\right)$ holds. Again by $\textsf{P2}$ we get
        $P\left(\sigma\meet\project{\sigma''}\right)$.

      \item[$\wedge$-intro]
        \[\infer{\sigma\into\vdash^{d}\Gamma,A_{0}\wedge
          A_{1}\gives\sigma'}{\sigma\into\vdash^{d}\Gamma,A_{i}\gives\sigma''\qquad
          \sigma''\into\vdash^{d}\Gamma,A_{1-i}\gives\sigma'}
        \]

        By applying the induction hypothesis to both premisses, we get
        $\sigma_{0},\sigma_{1}\in\Psi_{d}$ such that
        $\sigma''\obseq\sigma\meet\sigma_{0}$,
        $\sigma'\obseq\sigma''\meet\sigma_{1}$,
        $P\left(\sigma\meet\sigma_{0}\right)$,
        $P\left(\sigma''\meet\sigma_{1}\right)$, and
        $\vdash^{d}\Gamma,A_{i}\gives\sigma_{0}$ and
        $\vdash^{d}\Gamma,A_{1-i}\gives\sigma_{1}$ are derivable in {\DI}.
        Thus, $$\vdash^{d}\Gamma,A_{0}\wedge
        A_{1}\gives\sigma_{0}\meet\sigma_{1}$$ is derivable in {\DI}. Let us show
        that $\sigma_{0}\meet\sigma_{1}$ satisies each constraint:

        First, $\sigma_{0}\meet\sigma_{1}\in\Psi_{d}$ holds. Then, we have to
        show that $\sigma'\obseq\sigma\meet\sigma_{0}\meet\sigma_{1}$. Using
        $\textsf{D1}$,
        $\sigma''\meet\sigma_{1}\stronger\sigma''\stronger\sigma\meet\sigma_{0}$.
        Similarly, $\sigma''\meet\sigma_{1}\stronger\sigma_{1}$, and then
        $\sigma''\meet\sigma_{1}\stronger\sigma\meet\sigma_{0}\meet\sigma_{1}$.
        Thus, by transitivity,
        $$\sigma'\stronger\sigma\meet\sigma_{0}\meet\sigma_{1}$$ The other
        inequality holds by the same argument so that
        $\sigma'\obseq\sigma\meet\sigma_{0}\meet\sigma_{1}$. We have seen that
        $\sigma''\meet\sigma_{1}\stronger\sigma\meet\sigma_{0}\meet\sigma_{1}$.
        Hence, using $\textsf{P1}$ (left-to-right) and knowing
        $P\left(\sigma''\meet\sigma_{1}\right)$, we get
        $P\left(\sigma\meet\sigma_{0}\meet\sigma_{1}\right)$.

      \item[$\vee$-intro]
        \[\infer{\sig\into\vdash^{d}\Gamma,A\vee
          B\gives\sig'}{\sig\into\vdash^{d}\Gamma,A,B\gives\sig'}\]
        We conclude immediately with the induction hypothesis.

      \item[$\forall$-intro]
        \[\infer{\sig\into\vdash^{d}\Gamma,\forall x
          A\gives\sig'}{\sig\into\vdash^{\addFV{d}{\EVar{x}}}\Gamma,
          A\left[x:=\EVar{x}\right]\gives\sig'}\]
        We conclude immediately with the induction hypothesis.
      \end{itemize}
    \end{proof}
  \end{toappendix}
}

Notice that the statement for soundness  of Theorem~\ref{th:BB2sound}
is merely a
generalisation of axiom $\AXrp$ where the reference to $\BBci{d}$
and $\BBri{d}$ have respectively been replaced by derivability in
{\DI} and {\SDI}.

A natural statement for complete\-ness of {\SDI} \wrt {\DI} comes as
the symmetric generalisation of axiom $\AXpr$:

\begin{theorem}[Weak completeness of \SDI]\label{th:BB2wcomplete}
  If $\vdash^{d}\Gamma\gives \sigma'$ is derivable in {\DI}, then for
  all $\sigma\in\Psi_d$ such that $P(\sigma\wedge\sigma')$, there
  exists $\sigma''\in\Psi_d$ such that
  $\sigma''\obseq\sigma\meet\sigma'$ and
  $\sigma\into\vdash^{d}\Gamma\gives \sigma''$ is derivable in {\SDI}.
\end{theorem}
This statement can be proved, but it fails to capture an important
aspect of system {\SDI}: the order in which proof search treats
branches should not matter for completeness. But the above statement
concludes that there \emph{exists} a sequentialisation of branches
that leads to a complete proof tree in {\SDI}, so the proof-search
procedure should either guess it or investigate all possibilities.
\shortversion{We therefore proved~\cite{rouhling:hal-01107944} the
  stronger statement of completeness (below) whereby, }
\longversion{
  What we really want for completeness is a stronger statement whereby, }
\emph{for all} possible sequentialisations of branches, there exists a
complete proof tree. Therefore, when the proof-search procedure
decides to apply the branching rule, choosing which branch to complete
first can be treated as ``don't care non-determinism'' rather than
``don't know non-determinism'': if a particular choice proves
unsuccessful, there should be no need to explore the alternative
choice.

\longversion{
  \subsection{Proof of Completeness for \SDI}
  \label{sec:completenessSDI}

  A particular choice of sequentialisation can be represented by a
  black and white binary tree: the color of the nodes codes for the
  order in which branches are completed.
  \begin{definition}[Sequentialisation]\label{bwt}

    A \Index{sequentialisation} is an infinite binary tree whose
    nodes are labelled as either ``black'' or ``white''.

    We define the property, for a proof tree $\pi$ of system {\SDI}, to
    \Index[following a sequentialisation]{follow} a sequentialisation $r$, by
    induction on $\pi$:
    \begin{itemize}
    \item when the last rule of $\pi$ has no premiss, $\pi$ follows $r$;
    \item when the last rule of $\pi$ has one premiss, $\pi$ follows $r$ if its direct sub-proof tree does;
    \item when the last rule of $\pi$ is of the form
      \begin{footnotesize}
        \[\infer{\sig\into\vdash^{d}\Gamma,A_{0}\wedge A_{1}\gives\sig'}{
          \infer{\sig\into\vdash^{d}\Gamma,A_{i}\gives\sig''}{\pi_i}
          \qquad
          \infer{\sig''\into\vdash^{d}\Gamma,A_{1-i}\gives\sig'}{\pi_{1-i}}}\]
      \end{footnotesize}
      $\pi$ follows $r$ if
      \begin{itemize}
      \item $\pi_i$ follows the left direct sub-tree of $r$
      \item $\pi_{1-i}$ follows the right direct sub-tree of $r$
      \item either $i=0$ and the root of $r$ is ``white'', or $i=1$ and the root of $r$ is ``black''.
      \end{itemize}
    \end{itemize}
  \end{definition}

  This allows to quantify over all possible sequentialisations in the
  statement of the full completeness theorem:
}

\begin{toappendix}
    \begin{theorem}[Strong completeness of {\SDI}]\strut\label{th:BB2complete}

      If $\vdash^{d}\Gamma\gives \sigma'$ is derivable in {\DI}, then for
      all $\sigma\in\Psi_d$ such that $P(\sigma\wedge\sigma')$, and for
      all sequentialisations $r$ of branches, there exists
      $\sigma''\in\Psi_d$ such that $\sigma''\obseq\sigma\meet\sigma'$ and
      $\sigma\into\vdash^{d}\Gamma\gives \sigma''$ is derivable in {\SDI}
      with a proof tree that follows $r$.
    \end{theorem}
\end{toappendix}

\longversion{
  \begin{toappendix}[
    \begin{proof}See the proof in Appendix~\ref{sec:completenessSDI}.\end{proof}
    ]
  \begin{proof}
    By induction on the derivation of $\vdash^{d}\Gamma\gives\sig'$:
    \begin{itemize}

    \item[Theory]
      \[\infer[\mbox{$\BBc{\atmCtxt\Gamma}{\sigma'}{d}$}]{\vdash^{d}\Gamma\gives\sigma'}{}\]
      Let $\sigma\in\Psi_{d}$ such that $P\left(\sigma\meet\sigma'\right)$
      and $r$ be a sequentialisation. By $\AXpr$, there exists
      $\sigma''\in\Psi_{d}$ such that $\sigma''\obseq\sigma\meet\sigma'$
      and $\BBr{\sigma}{\atmCtxt\Gamma}{\sigma''}{l}$. Thus,
      $\sigma\into\vdash^{d}\Gamma\gives\sigma''$ is derivable in {\SDI}
      using the theory rule. The proof tree follows $r$, for it has no premiss.

    \item[$\exists$-intro]
      \[
        \infer{\vdash^{d}\Gamma,\exists
        xA\gives\project{\sigma'}}{\vdash^{\addFV{d}{\XVar{X}}}\Gamma,
        A\left[x:=\XVar{X}\right],\exists xA\gives\sigma'}
      \]
      By induction hypothesis, for all $\sigma\in\Psi_{\addFV{d}{\XVar{X}}}$
      such that $P\left(\sigma\meet\sigma'\right)$, for all sequentialisations $r$, there exists $\sigma_{0}\in\Psi_{\addFV{d}{\XVar{X}}}$ such
      that $\sigma_{0}\obseq\sigma\meet\sigma'$ and
      \[
        \sigma\into\vdash^{\addFV{d}{\XVar{X}}}\Gamma,
          A\left[x:=\XVar{X}\right],\exists xA\gives\sigma_{0}
      \]
      is derivable in {\SDI} with a proof tree following $r$. Let
      $\sigma\in\Psi_{d}$ such that
      $P\left(\sigma\meet\project{\sigma'}\right)$ and $r$ be a sequentialisation. If we find $\sigma''\in\Psi_{\addFV{d}{\XVar{X}}}$ such that
      $\sigma''\obseq\inject{\sigma}\meet\sigma'$ and
      \[
        \inject{\sigma}\into\vdash^{\addFV{d}{\XVar{X}}}\Gamma,
        A\left[x:=\XVar{X}\right],\exists xA\gives\sigma''
      \]
      is derivable in {\SDI} with a proof tree following $r$, then we can
      conclude by using $\textsf{D2}$ and Definition~\ref{bwt}.  Since
      $\inject{\sigma}\in\Psi_{\addFV{d}{\XVar{X}}}$, we just need to show that
      $P\left(\inject{\sigma}\meet\sigma'\right)$ holds and to apply the
      induction hypothesis.  By $\textsf{D2}$,
      $\project{(\inject{\sigma}\meet\sigma')}\obseq\sigma\meet\project{\sigma'}$.
      Hence, from the assumption $P\left(\sigma\meet\project{\sigma'}\right)$
      we derive $P\left(\project{(\inject{\sigma}\meet\sigma')}\right)$ with
      $\textsf{P2}$, and then by $\textsf{P1}$ (right-to-left) we conclude
      $P\left(\inject{\sigma}\meet\sigma'\right)$.

    \item[$\wedge$-intro]
      \[
        \infer{\vdash^{d}\Gamma,A_{0}\wedge
        A_{1}\gives\sig_{0}\meet\sig_{1}}{\vdash^{d}\Gamma,A_{0}\gives\sig_{0}
        \qquad \vdash^{d}\Gamma,A_{1}\gives\sig_{1}}
      \]
      Let $\sigma\in\Psi_{d}$ such that
      $P\left(\sigma\meet\sigma_{0}\meet\sigma_{1}\right)$ and $r$ be a sequentialisation. Without loss of generality, we can assume that the root
      of $r$ is white. $\textsf{D1}$ gives us
      $\sigma\meet\sigma_{0}\meet\sigma_{1}\stronger\sigma\meet\sigma_{0}$.
      Then, with $\textsf{P1}$ (left-to-right), we get
      $P\left(\sigma\meet\sigma_{0}\right)$.  Hence, we can apply the induction
      hypothesis on the left subtrees and on $\sigma$.  So, there exists
      $\sigma'\in\Psi_{d}$ such that \[\sigma'\obseq\sigma\meet\sigma_{0}\text{
      and }\sigma\into\vdash^{d}\Gamma,A_{0}\gives\sigma'\] is derivable in
      {\SDI} with a proof tree following the left subtree of $r$.

      Using $\textsf{D1}$, we easily obtain
      $\sigma\meet\sigma_{0}\meet\sigma_{2}\stronger\sigma'\meet\sigma_{1}$ and
      then with $\textsf{P1}$ (left-to-right) we get
      $P\left(\sigma'\meet\sigma_{1}\right)$\footnote{Recall the manipulations
      done in the $\wedge$-intro part of the proof of Theorem~\ref{th:BB2sound};
      those are similar.}. Thus, by applying the induction
      hypothesis to the right subtrees and to $\sigma'$, we get
      $\sigma''\in\Psi_{d}$ such that
      $\sigma''\obseq\sigma'\meet\sigma_{1}$ and
      \[\sigma'\into\vdash^{d}\Gamma,A_{1}\gives\sigma''\] is derivable in
      {\SDI} with a proof tree following the right subtree of $r$.

      Thus, $$\sigma\into\vdash^{d}\Gamma,A_{0}\wedge
      A_{1}\gives\sigma''$$ is derivable in {\SDI} with a proof tree
      following $r$. There remains to show that
      $\sigma''\obseq\sigma\meet\sigma_{0}\meet\sigma_{1}$, knowing
      $\sigma''\obseq\sigma'\meet\sigma_{1}$ and
      $\sigma'\obseq\sigma\meet\sigma_{0}$. This kind of proof has been done in
      the $\wedge$-intro part of the proof of Theorem~\ref{th:BB2sound}.

    \item[$\vee$-intro]
      \[\infer{\vdash^{d}\Gamma,A\vee B\gives\sig'}{\vdash^{d}\Gamma,A,B\gives\sig'}\]
      We conclude immediately with the induction hypothesis.

    \item[$\forall$-intro]
      \[
        \infer{\vdash^{d}\Gamma,\forall x A\gives\sig'}{\vdash^{\addFV{d}{\EVar{x}}}
        \Gamma,A\left[x:=\EVar{x}\right]\gives\sig'}
      \]
      We conclude immediately with the induction hypothesis.

    \end{itemize}
  \end{proof}
\end{toappendix}
}

\section{Relating {\SDI} to {\LKone}}
\label{sec:relating}

Now we combine the two steps: from {\LKone} to {\DI} and from {\DI} to
{\SDI}, so as to relate {\SDI} to {\LKone}.
For this we aggregate (and consequently simplify) the axioms that we used for
the first step with those that we used for the second step.

\begin{definition}[Compatibility-based pre-order]\strut\label{def:comppo}
  Assume we have a family of compatibility relations $\compat$ for a constraint
  structure $(\Psi_d)_d$.  We define the following pre-order on each $\Psi_d$:
  \[
    \forall\sig,\sig'\in\Psi_d,\ \sig\stronger[\compat]\sig' \Leftrightarrow
    \left\{\rho\in \Inst{d}\mid\rho\compat\sig\right\}\subseteq
    \left\{\rho\in \Inst{d}\mid\rho\compat\sig'\right\}
  \]
  and let $\obseq[\compat]$ denote the symmetric closure of $\stronger[\compat]$.
\end{definition}

We now assume that we have a lift constraint structure and a
constraint-refining predicate $(\BBri{d})_d$ used to define {\SDI}, and the
existence of
\begin{itemize}
  \item a binary operator $\meet$
  \item a compatibility relation $\compat$ that distributes over
    $\meet$\hfill({\AXproj} and {\AXmeet} in Fig.~\ref{fig:summary})
  \item a binding operator for $\compat$\hfill(\AXwitness{} in
    Fig.~\ref{fig:summary})
  \item a constraint-producing predicate $(\BBci{d})_d$ that relates to
    $\BBi$\hfill($\AXpg$ in Fig.~\ref{fig:summary})
  \item a predicate $P$
\end{itemize}
satisfying the axioms of Fig.~\ref{fig:summary}.
\longversion{
  \subsection{Decency}
  \label{sec:decency}
}
\shortversion{These entail decency~\cite{rouhling:hal-01107944}:}
\longversion{
  The axioms of Fig.~\ref{fig:summary} entail decency (see
  Appendix~\ref{sec:decency}):}
\begin{toappendix}
  \begin{lemma}\label{lem:decent}
    Given the axioms of Fig.~\ref{fig:summary}, $(\stronger[\compat],\meet,P)$ is decent.
  \end{lemma}
\end{toappendix}

\begin{toappendix}[]
  \begin{proof}
    First, notice that Axiom $\AXmeet$ (together with
    Definition~\ref{def:comppo}) makes $\meet$ implement a set-theoretic
    intersection in the following sense:
    \[
      \left\{\rho\in \Inst{d}\mid\rho\compat(\sig\meet\sig')\right\} =
      \left\{\rho\in \Inst{d}\mid\rho\compat\sig\right\}\cap\left\{\rho\in
      \Inst{d}\mid\rho\compat\sig'\right\}
    \]
    \begin{itemize}
      \item[\textsf{D1}] This is a direct consequence of the above remark.
      \item[\textsf{D2}] Assume
        $\sigma''\obseq[\compat]\inject\sigma\meet\sigma'$. We prove
        $\project{\sigma''}\obseq[\compat]\sigma\meet\project{\sigma'}$.

        Take $\rho\compat\project{\sigma''}$. By Axiom $\AXwitness$ we have
        $(\addInst{\XVar{X}}{\bind{\sigma''}{(\rho)}}{\rho})\compat\sigma''$. By
        Definition~\ref{def:comppo} and the above remark we have
        $(\addInst{\XVar{X}}{\bind{\sigma''}{(\rho)}}{\rho})\compat\inject\sigma$ and
        $(\addInst{\XVar{X}}{\bind{\sigma''}{(\rho)}}{\rho})\compat\sigma'$. By Axiom
        $\AXlift$ (left-to-right) we have $\rho\compat\sigma$.  By Axiom
        $\AXproj$ we also have $\rho\compat\project{\sigma'}$. Hence we
        have $\rho\compat\project{\sigma'}$. So
        $\rho\compat(\sigma\meet\project{\sigma'})$ by the above remark.
        Hence,
        $\project{\sigma''}\stronger[\compat]\sigma\meet\project{\sigma'}$.

        Conversely, take $\rho\compat(\sigma\meet\project{\sigma'})$. By the
        above remark, $\rho\compat\sigma$ and $\rho\compat\project{\sigma'}$.
        By Axiom $\AXwitness$ we have
        $(\addInst{\XVar{X}}{\bind{\sigma'}{(\rho)}}{\rho})\compat\sigma'$, and by Axiom
        $\AXlift$ (right-to-left) we have
        $(\addInst{\XVar{X}}{\bind{\sigma'}{(\rho)}}{\rho})\compat\inject\sigma$. By the above
        remark we have
        $(\addInst{\XVar{X}}{\bind{\sigma'}{(\rho)}}{\rho})\compat(\inject\sigma\meet\sigma')$
        and by Definition~\ref{def:comppo} we have
        $(\addInst{\XVar{X}}{\bind{\sigma'}}{(\rho)}{\rho})\compat\sigma''$. By Axiom
        $\AXproj$ we have $\rho\compat\project{\sigma''}$.  Hence,
        $\sigma\meet\project{\sigma'}\stronger[\compat]\project{\sigma''}$.
    \end{itemize}
  \end{proof}
\end{toappendix}

Hence, we have soundness and completeness of {\SDI} \wrt {\LKone} on the empty
domain, as a straightforward consequence of
Corollary~\ref{cor:SDIsoundcomplempty} and Theorems~\ref{th:BB2sound} and~\ref{th:BB2complete}:
%% \begin{theorem}[Completeness of {\SDI}]

%%   If $\vdash_{n}\rho(\Gamma)$ is derivable in {\LKone}, then there exists $\sigma_1\in\Psi_l$ such that, for all $\sigma\in\Psi_l$ with $P(\sigma\wedge\sigma_1)$ and $\rho\compat\sigma$, and for all sequentialisation $r$, there exists $\sigma'\in\Psi_l$ such that $\sigma'\obseq\sigma\meet\sigma_1$, $\rho\compat\sigma'$, and $\sigma\into\vdash_{n}^{l}\Gamma\gives \sigma'$ is derivable in {\SDI} with a proof tree that follows $r$.
%% \end{theorem}

\begin{theorem}[Soundness and completeness on the empty domain]\strut

  If $\sigma\into\vdash^{\dinit}\Gamma\gives \sigma'$ is derivable in
  {\SDI} and $\emptyInst\compat\sigma'$, then $\vdash\Gamma$ is derivable
  in {\LKone}.
  In particular when $P$ is the predicate ``being satisfiable'', if
  \mbox{$\sigma\into\vdash^{\dinit}\Gamma\gives \sigma'$} is derivable in
  {\SDI}, then $\vdash\Gamma$ is derivable in {\LKone}.

  Assume $P$ is always true or is ``being satisfiable''.  If
  $\vdash\Gamma$ is derivable in {\LKone}, then for all
  $\sigma\in\Psi_{\dinit}$ such that $\emptyInst\compat\sigma$ and for
  all sequentialisations $r$, there exists $\sigma'\in\Psi_{\dinit}$
  such that $\emptyInst\compat\sigma'$ and
  $\sigma\into\vdash^{\dinit}\Gamma\gives \sigma'$ is derivable in
  {\SDI} with a proof tree that follows $r$.
\end{theorem}

\begin{remark}[Soundness of {\SDI}]
  Soundness of \SDI{} on an arbitrary domain is a
  direct consequence of Theorem~\ref{th:BB1sound} and
  Theorem~\ref{th:BB2sound}: If $\sigma\into\vdash^{d}\Gamma\gives
  \sigma'$ is derivable in {\SDI}, then $P(\sigma')$ holds and for all
  $\rho\compat\sigma'$, $\vdash\rho(\Gamma)$ is derivable in
  {\LKone}. For the sake of brevity, we omit the general statement of
  completeness on an arbitrary domain, which is quite long to write.
\end{remark}

As we shall see in Sect.~\ref{sec:implem}, it is useful to have a ``top
element'' $\top$ in $\Psi_{\dinit}$ with $\emptyInst\compat\top$, which we feed
to a proof-search procedure based on {\SDI}, as the initial input constraint
$\sigma$ mentioned in the soundness and completeness theorems.

\begin{figure}[h]
  \[
  \begin{array}{|c|}
    \hline
    \begin{array}{l@{\qquad}lll}      
      \AXproj
      &\forall\sigma\in\Psi_{\addFV d{\XVar X}},
      &\forall t \forall \rho,
      & \left(\addInst{\XVar{X}}{t}{\rho}\right)\compat\sig\Rightarrow \rho\compat\project\sig
      % \\
      % \AXproj[!]
      % &\forall\sigma\in\Psi_{\addFV d{\EVar x}},
      % &\forall \rho,
      % & \rho\compat\sig\Leftrightarrow \rho\compat\project\sig
      \\
      \AXwitness
      &\forall\sigma\in\Psi_{\addFV d{\XVar X}},
      &\forall\rho,
      &\rho\compat\project\sig\Rightarrow\left(\addInst{\XVar{X}}{\bind\sigma\left(\rho\right)}{\rho}\right)\compat\sig
      \\
      \AXmeet
      &\forall \sig \sig'\in\Psi_d,
      &\forall \rho,
      &
      \left\{
        \begin{array}{l}
          \rho\compat\sig\\
          \rho\compat\sig'
        \end{array}
      \right.
      \Leftrightarrow\rho\compat\left(\sig\meet\sig'\right)
      \\
      \AXpg
      &
      \multicolumn{3}{c}{\forall l,\forall \mathcal A,\qquad
        \left\{\rho\mid\;
          \BB{\rho\left(\mathcal A\right)}\right\}=\bigcup\limits_{\{\sig\mid\;\BBc{\mathcal A}{\sig}{l}\}}\left\{\rho\mid\rho\compat\sig\right\}
      }
    \end{array}\\
    \hline\hline
    \begin{array}{l@{\qquad}lll}
      \AXlift
      &\forall \sigma\in\Psi_d,\forall\sigma'\in\Psi_{\addFV d{\XVar X}},
      &\forall \rho,
      &(\addInst{\XVar{X}}{\bind{\sigma'}{(\rho)}}{\rho})\compat\inject\sigma\ \Leftrightarrow\ \rho\compat\sigma
      \\
      \AXp[1]
      &\forall\sigma\in\Psi_{\addFV d{\XVar X}},
      && P(\sig)\Leftrightarrow P(\project\sig)
      \\
      \AXp[2]
      &\forall\sigma \sigma'\in\Psi_d,
      &&\left\{
        \begin{array}l
          P(\sig)\\
          \sig\stronger[\compat]\sig'
        \end{array}\right.\Rightarrow P(\sig')
      \\
      \AXrp
      &
      \multicolumn{3}{c}{
        \forall d,\forall \mathcal A,\forall \sigma,\sigma'\in\Psi_d,\quad \BBr{\sig}{\mathcal A}{\sig'}{d}
        \Rightarrow\ \exists \sigma''\in\Psi_d,
        \left\{
          \begin{array}l
            \sigma'\obseq[\compat]\sig\meet\sig''\\
            P(\sig\meet\sig'')\\
            \BBc{\mathcal A}{\sig''}{d}
          \end{array}
        \right.}
      \\
      \AXpr
      &
      \multicolumn{3}{c}{
        \forall d,\forall \mathcal A,\forall \sigma,\sigma'\in\Psi_d,\quad
        \left\{
          \begin{array}l
            P(\sig\meet\sig')\\
            \BBc{\mathcal A}{\sig'}{d}
          \end{array}
        \right.
        \Rightarrow\ \exists \sigma''\in\Psi_d,
        \left\{
          \begin{array}l
            \sigma''\obseq[\compat]\sig\meet\sig'\\
            \BBr{\sig}{\mathcal A}{\sig''}{d}
          \end{array}
        \right.}
    \end{array}\\
  \hline
  \end{array}
  \]
  \caption{Full Axiomatisation}
  \label{fig:summary}
\end{figure}

\medskip
\section{Implementation}
\label{sec:implem}

{\Psyche} is a platform for proof search, where a \emph{kernel} offers
an API for programming various search strategies as \emph{plugins},
while guaranteeing the correctness of the search
output~\cite{GLPsyche13}. Its architecture extensively uses OCaml's
system of

\begin{wraptable}r{190pt}\vspace{-5pt}
\begin{lstlisting}[frame=single,title=Theory component signature in \Psyche\ 2.0,basicstyle=\ttfamily\footnotesize,columns=fullflexible,captionpos=b,label=list:theorysig]
module type Theory = sig
 module Constraint: sig
   type t
   val topconstraint:t
   val proj : t -> t
   val lift : t -> t
   val meet : t -> t -> t option
   ...
 end
 val consistency :
  ASet.t -> (ASet.t,Constraint.t) stream
end
\end{lstlisting}
\end{wraptable}
% based
% on the \emph{focused} sequent calculus developed
% in~\cite{FarooquePhD}\longversion{ and presented in
% Appendix~\ref{focused}}.
\noindent modules and functors. In order to modularly support
theory-specific reasoning (in presence of quantifiers), the
axiomatisation proposed in the previous sections was used to identify
the signature and the specifications of \emph{theory components}.
In version 2.0 of \Psyche~\cite{Psyche}, the kernel implements (the
\emph{focused} version of) System \SDI, and a theory component is
required to provide  the implementation of the concepts\shortversion{\clearpage\noindent} developed in
the previous sections, as shown in the module type above.  It provides
a lift constraint structure in the form of a module
{\footnotesize\verb=Constraint=}, with a type for constraints, the
projection and lift maps, as well as a top constraint (always
satisfied) with which proof search will start. We also require a meet
operation: While the theory of complete proofs in {\SDI}
does not need it, the meet operation is useful when implementing a
backtracking proof-search procedure: imagine a proof tree has been
completed for some sequent $\mathcal S$, with input constraint
$\sigma_0$ and output constraint $\sigma_1$; at some point the
procedure may have to search again for a proof of $\mathcal S$ but
with a different input constraint $\sigma_0'$. We can check whether
the first proof can be re-used by simply checking whether
$\sigma_0'\wedge\sigma_1$ is satisfiable. The
{\footnotesize\verb=meet=} function should output
{\footnotesize\verb=None=} if the meet of the two input constraints is
not satisfiable, and {\footnotesize\verb=Some sigma=} if the
satisfiable meet is {\footnotesize\verb=sigma=}.

Finally, the function that is called at the leaves of proof trees is
{\footnotesize\verb=consistency=}, which implements the constraint-refining
predicate; {\footnotesize\verb=ASet.t=} is the type for sets of literals with
meta-variables and the function returns a stream: providing an input
constraint triggers computation and pops the next element of the
stream if it exists. It is a pair made of an output constraint
and a subset of the input set of literals. The latter indicates which literals
of the input have been used to close the branch, which is useful information
for \emph{lemma learning} (see \eg \cite{GLPsyche13}).

While our axiomatisation immediately yields the specification for
theory components, it does not provide instances and so far, the only
(non-ground) instance implemented in \Psyche\ is that of pure
first-order logic (based on unification). \longversion{Our next steps will be to
implement other theories, such as those admitting quantifier
elimination like linear rational and integer arithmetic.
Finally, as we aim at making {\Psyche} available for deductive
verification, we have started to work on its integration to the ``herd
of provers'' of the Why3 platform~\cite{boogie11why3}.}
\section{Related Works and Further Work}
\label{RelatedWorks}
\label{Conclusion}

The sequent calculi developed in this paper for theory reasoning in
presence of quantifiers, are akin to the free variable tableaux
of~\cite{beckert99handbook} for total theory reasoning.
\longversion{Particularly
System \SDI\ where a branch closure affects the remaining open
branches in an asymmetric treatment.

}
But they use abstract constraints, instead of substitutions, and
our foreground reasoner is able to propagate them across
branches while being ignorant of their nature. This allows new
theories to be treated by the framework, such as those satisfying
quantifier elimination, like linear arithmetic.  In this particular
case, the asymmetric treatment of \SDI\ formalises an improvement, in
the view of an effective implementation, over System
PresPred${}^C_S$~\cite{princess08} for LIA.  A novel point of our
paper is to show that the propagation of substitutions in
tableaux and the propagation of linear arithmetic constraints
follow the same pattern, by describing them as two instances of an
abstract constraint propagation mechanism.

Constraints have been integrated to various tableaux calculi:
In the nomenclature proposed in Giese and H\"ahnle’s
survey~\cite{GieseHaehnle03}, our approach is closest to
\emph{constrained formula tableaux} or \emph{constrained branch
  tableaux} which propagate constraints between branches (rather than
\emph{constrained tableaux} which have a global management of
constraints). But the \emph{tableaux} calculi cited
by~\cite{GieseHaehnle03} in these categories are for specific theories
and logics (pure classical logic, equality, linear temporal logic or
bunched implications), in contrast to our generic approach.

When classes of theories are generically integrated to automated
reasoning with the use of constraints, as for the Model Evolution
Calculus~\cite{BaumgartnerT11}, these are usually described as
first-order formulae over a particular theory's signature (as it is
the case in~\cite{BFT2008MELIA,princess08} for LIA). Our abstract
data-structures for constraints could be viewed as the semantic
counter-part of such a syntactic representation, whose atomic
construction steps are costless but which may incur expensive
satisfiability checks by the background reasoner. Our semantic view of
constraints, as shown in Section~\ref{sec:implem}, more directly
supports theory-tuned implementations where \eg the meet and
projection operations involve computation. Our specifications for
theory-specific computation also seems less demanding than deciding
the satisfiability of any constraint made of atoms (over the theory's
signature), conjunction, negation, and existential
quantification~\cite{BaumgartnerT11}.

\longversion{The semantic approach features the following
  computational trade-off: when each constraint covers a large pack of
  possible instantiations, the streams produced at the leaves of our
  proof trees may enumerate the set of solutions at a fast pace, but
  operations on constraints may become costly (\eg quantifier
  elimination); conversely, simple constraints (\eg when a constraint
  is a single instantiation) may have more efficient operations for
  them, but as each of them covers fewer instantiations, the
  enumeration of solutions is slower.}

The semantic approach to constraints was explored by a rich
literature in (Concurrent) Constraint Programming~\cite{Saraswat91},
but the applicability of constraint systems to programming usually
leads to more demanding axioms as well (requiring \eg complete
lattices) and to a global management of constraints (with a global
store that is reminiscent of \emph{constrained tableaux}). Our local
management of constraints allows for more subtle backtracking
strategies in proof search, undoing some steps in one branch while
sticking to some more recent decisions that have been made in a
different branch.

\longversion{Although constraint-based unification algorithms or universal
unification algorithms could provide theory reasoners for our
framework, our paper is \emph{not} about using constraints to specify
unification modulo a theory. This approach, in the nomenclature of theory
reasoning~\cite{baumgartner1992unified}, pertains to the class of
\emph{term level interactions} between a foreground reasoner and a
background theory reasoner. The interaction we describe remains at the
\emph{literal level}. It also differs from \emph{Deduction
  Modulo}~\cite{DowekHK03}, in which theory reasoners are rewrite
systems that can only act upon one literal of a sequent at a time, and
therefore cannot capture reasoners (such as a simplex algorithm) that
derive \eg a theory inconsistency from an input collection of
literals, as used here or in SMT-solving.}

In the case of ground theory reasoning, the field of SMT-solving has
evolved powerful techniques for combining theories (see \eg the
unifying approach of~\cite{Shankar04techreport}). A natural question
is whether similar techniques can be developed in presence of
quantifiers, combining constraint-producing or constraint-refining
procedures. We did not provide such techniques here, but we believe
our modular and abstract approach could be a first step towards that
end, with our axiomatisation identifying what properties should be
sought when engineering such techniques, \ie serving as a correctness
criterion.

Finally, SMT-solvers usually adopt a heuristic approach for handling
quantifiers, often involving incomplete mechanisms, with slimmer
theoretical foundations than for their ground reasoning core. A
notable exception is a formalisation of \emph{triggers} mechanisms by
Dross et al.~\cite{DrossCKP12}, which we hope to view as particular
instances of our constraint systems. Moreover, the way in which
triggers control the breaking of quantifiers appears as the kind of
structured proof-search mechanisms that \Psyche\ can specify (based on
focusing).% We

\textbf{Acknowledgements.}  {\small This research was supported by ANR
  projects PSI and ALCOCLAN, as well as by DARPA under agreement
  number FA8750-12-C-0284. The views and conclusions contained herein
  are those of the authors and should not be interpreted as
  necessarily representing the official policies or endorsements,
  either expressed or implied, of DARPA, or the U.S.\ Government.}

\def\homedir{\textasciitilde}

\ifdefined\url
\else
\def\url[#1]#2{\texttt{#2}}
\fi

\let\oldurl\url

\makeatletter
\ifdefined\href
\def\myurl{\@ifnextchar[\@myurlwith\@myurlwithout}
\long\def\@myurlwith[#1]#2{\mbox{\href{#2}{#1}}}
\long\def\@myurlwithout#1{\mbox{\href{#1}{#1}}}
\else
\def\myurl{\@ifnextchar[\@myurlwith\@myurlwithout}
\long\def\@myurlwith[#1]#2{\mbox{\oldurl[#1]{#2}}}
\long\def\@myurlwithout#1{\mbox{\oldurl[]{#1}}}
\fi
\makeatother

\def\url{\myurl}

% COLOR FORMAT
% \newcommand\emphasizeformat[1]{{\color{BrickRed}{{\em #1}}}}
% \newcommand\authorformat[1]{{\textcolor[rgb]{0,0,1}{#1}}}

% SMALL CAPITAL FORMAT
% \newcommand\emphasizeformat[1]{{\em #1}}
% \newcommand\authorformat[1]{{\sc #1}}

% NORMAL FORMAT
\newcommand\emphasizeformat[1]{\emph{#1}}
\newcommand\authorformat[1]{{#1}}
% DISPLAY MONTH?

\newcommand\monthdisplay[1]{\unskip}

\bibliographystyle{abbrv}
\bibliography{Common/abbrev-short,Common/Main,Common/crossrefs}

\begin{thebibliography}{10}

\bibitem{BFT2008MELIA}
P.~Baumgartner, A.~Fuchs, and C.~Tinelli.
\newblock {ME(LIA)} -- {M}odel {E}volution {W}ith {L}inear {I}nteger
  {A}rithmetic {C}onstraints.
\newblock In I.~Cervesato, H.~Veith, and A.~Voronkov, editors, {\em Proc.{} of
  the the 15th Int.{} Conf.{} on Logic for Programming Artificial Intelligence
  and Reasoning (LPAR'08)}, volume 5330 of {\em LNCS}, pages 258--273.
  Springer-Verlag, Nov. 2008.

\bibitem{baumgartner1992unified}
P.~Baumgartner, U.~Furbach, and U.~Petermann.
\newblock A unified approach to theory reasoning.
\newblock Technical report, Inst. f{\"u}r Informatik, Univ., 1992.

\bibitem{BaumgartnerT11}
P.~Baumgartner and C.~Tinelli.
\newblock Model evolution with equality modulo built-in theories.
\newblock In N.~Bj{\o}rner and V.~Sofronie-Stokkermans, editors, {\em Proc.{}
  of the 23rd Int.{} Conf.{} on Automated Deduction (CADE'11)}, volume 6803 of
  {\em LNCS}, pages 85--100. Springer-Verlag, July 2011.

\bibitem{Beckert98}
B.~Beckert.
\newblock Chapter 8: Rigid {$E$}-unification.
\newblock In W.~Bibel and P.~H. Schmitt, editors, {\em Automated Deduction -- A
  Basis for Applications}, volume I: Foundations. Calculi and Methods, pages
  265--289. Kluwer Academic Publishers, 1998.

\bibitem{beckert99handbook}
B.~Beckert.
\newblock Equality and other theories.
\newblock In {\em Handbook of Tableau Methods}, pages 197--254. Kluwer Academic
  Publishers, 1999.

\bibitem{boogie11why3}
F.~Bobot, J.-C. Filli\^atre, C.~March\'e, and A.~Paskevich.
\newblock Why3: Shepherd your herd of provers.
\newblock In {\em Boogie 2011: First International Workshop on Intermediate
  Verification Languages}, pages 53--64, Aug. 2011.

\bibitem{DowekHK03}
G.~Dowek, T.~Hardin, and C.~Kirchner.
\newblock Theorem proving modulo.
\newblock {\em J. of Automated Reasoning}, 31(1):33--72, 2003.

\bibitem{DrossCKP12}
C.~Dross, S.~Conchon, J.~Kanig, and A.~Paskevich.
\newblock Reasoning with triggers.
\newblock In P.~Fontaine and A.~Goel, editors, {\em 10th Int.{} Work.{} on
  Satisfiability Modulo Theories, {SMT} 2012}, volume~20 of {\em EPiC Series},
  pages 22--31. EasyChair, June 2012.

\bibitem{FarooquePhD}
M.~Farooque.
\newblock {\em Automated reasoning techniques as proof-search in sequent
  calculus}.
\newblock PhD thesis, Ecole Polytechnique, 2013.

\bibitem{Shankar04techreport}
H.~Ganzinger, H.~RueB, and N.~Shankar.
\newblock Modularity and refinement in inference systems.
\newblock Technical Report SRI-CSL-04-02, SRI, 2004.

\bibitem{Giese00}
M.~Giese.
\newblock Proof search without backtracking using instance streams, position
  paper.
\newblock In P.~Baumgartner and H.~Zhang, editors, {\em 3rd Int.{} Work.{} on
  First-Order Theorem Proving (FTP), St.~Andrews, Scotland, TR 5/2000 Univ.\ of
  Koblenz}, pages 227--228, 2000.

\bibitem{GieseHaehnle03}
M.~Giese and R.~H{\"a}hnle.
\newblock Tableaux + constraints.
\newblock In M.~C. Mayer and F.~Pirri, editors, {\em Proc.{} of the 16th Int.{}
  Conf.{} on Automated Reasoning with Analytic Tableaux and Related Methods
  (Tableaux'03)}, volume 2796 of {\em LNCS}, pages 37--42. Springer-Verlag,
  Sept. 2003.

\bibitem{GLPsyche13}
S.~Graham-Lengrand.
\newblock {Psyche}: a proof-search engine based on sequent calculus with an
  {LCF}-style architecture.
\newblock In D.~Galmiche and D.~Larchey-Wendling, editors, {\em Proc.{} of the
  22nd Int.{} Conf.{} on Automated Reasoning with Analytic Tableaux and Related
  Methods (Tableaux'13)}, volume 8123 of {\em LNCS}, pages 149--156.
  Springer-Verlag, Sept. 2013.

\bibitem{Nieuwenhuis06}
R.~Nieuwenhuis, A.~Oliveras, and C.~Tinelli.
\newblock Solving {SAT} and {SAT Modulo Theories}: From an abstract
  {Davis}--{Putnam}--{Logemann}--{Loveland} procedure to {DPLL({\it T})}.
\newblock {\em J. of the ACM Press}, 53(6):937--977, 2006.

\bibitem{Psyche}
Psyche: the {Proof-Search factorY for Collaborative HEuristics}.

\bibitem{princess08}
P.~R{\"u}mmer.
\newblock A constraint sequent calculus for first-order logic with linear
  integer arithmetic.
\newblock In I.~Cervesato, H.~Veith, and A.~Voronkov, editors, {\em Proc.{} of
  the the 15th Int.{} Conf.{} on Logic for Programming Artificial Intelligence
  and Reasoning (LPAR'08)}, volume 5330 of {\em LNCS}, pages 274--289.
  Springer-Verlag, Nov. 2008.

\bibitem{Saraswat91}
V.~A. Saraswat, M.~Rinard, and P.~Panangaden.
\newblock The semantic foundations of concurrent constraint programming.
\newblock In D.~S. Wise, editor, {\em 18th Annual ACM Symp.{} on Principles of
  Programming Languages (POPL'91)}, pages 333--352. ACM Press, Jan. 1991.

\bibitem{Stickel85}
M.~E. Stickel.
\newblock Automated deduction by theory resolution.
\newblock {\em J. of Automated Reasoning}, 1(4):333--355, 1985.

\end{thebibliography}

\longversion{
\newpage
\appendix

\section{Focused Sequent Calculus}
\label{focused}

In this section we give (a fragment of) the focused sequent
calculus from~\cite{FarooquePhD}, called \LKThp, on which
the \Psyche\ implementation is based.

\begin{figure}[!h]
  \[
  \begin{array}{|c|}
    \upline
    \textsf{Synchronous rules}
    \hfill\strut\\[3pt]
    %% \infers[(\andP)]{\DerPos{\Gamma}{A\andP B}{}}
    %% {\DerPos{\Gamma}{A} {} \qquad \DerPos{\Gamma}{B}{}}
    %% \qquad
    \infers[(\orP)]{\DerPos{\Gamma}{A_1\orP A_2}{}}
    {\DerPos{\Gamma}{A_i}{}}
       \qquad
    \infers[(\EX)]{\DerPos{\Gamma}{\EX x A}{}}
    {\DerPos{\Gamma}{A[x:=t]}{}}
    \\[15pt]
\infers[{(\Init[1])}]{\DerPos{\Gamma  }{l}{} }[l \mbox{ is positive}]
        {\atmCtxt\Gam,\non l\models}    \qquad
    \infers[(\Release)]{\DerPos {\Gam} {N} {} }[N \mbox{ is negative}]
    {\DerNeg {\Gam} {N} {} }\\
    \midline
    \textsf{Asynchronous rules}
    \hfill\strut\\[3pt]
    \infers[(\andN)]{\DerNeg{\Gamma}{A\andN B,\Delta} {} }
    {\DerNeg{\Gamma}{A,\Delta} {}
      \qquad \DerNeg{\Gamma}{B,\Delta} {} }
    %% \qquad
    %% \infers[(\orN)]{\DerNeg {\Gamma} {A_1\orN A_2,\Delta} {} }
    %% {\DerNeg {\Gamma} {A_1,A_2,\Delta} {} }

   \qquad
       \infers[(\FA)]{\DerNeg{\Gamma}{(\FA x A),\Delta}{} }[x\notin\FV{\Gam,\Delta}]
    {\DerNeg {\Gamma} {A,\Delta}{} }\\\\
    \infers[({\Store})]%
      {\DerNeg \Gam {A,\Del} {} }
      [\begin{array}l%
        A\mbox{ is a literal}\\
        \mbox{or is positive}\\
      \end{array}]
    {\DerNeg {\Gam,\non A} {\Del} {} {}}
    \\\midline
      \textsf{Structural rules}
    \hfill\strut\\[3pt]
    \infers[(\Select)]
    {\DerNeg {\Gam,\non P} {}{} } [\mbox{$P$ is positive}]
    {\DerPos {\Gam,\non P} {P} {} }
    \qquad
    \infers[({\Init[2]})]{\DerNeg {\Gam} {}{} }{\atmCtxt\Gam\models}
    \\\downline
  \end{array}
  \]

  \caption{System \LKThp}
  \label{fig:LKThp}
\end{figure}

\section{Full Proofs}

\label{appendix}
\gettoappendixwr {th:BB1sound}
\gettoappendix {th:BB1soundproof}
\gettoappendixwr {cor:soundnesscanonical}
\gettoappendixwr {th:BB2sound}
\gettoappendix {th:BB2soundproof}
\gettoappendixwr {th:BB2complete}
\gettoappendix {th:BB2completeproof}
\gettoappendixwr {lem:decent}
\gettoappendix {lem:decentproof}

}

\end{document}